\documentclass[fleqn,usenatbib]{mnras}

\usepackage[T1]{fontenc}

\DeclareRobustCommand{\VAN}[3]{#2}
\let\VANthebibliography\thebibliography
\def\thebibliography{\DeclareRobustCommand{\VAN}[3]{##3}\VANthebibliography}

\usepackage{graphicx}	
\usepackage{amsmath}	
\usepackage{amssymb}	
\usepackage{bm}
\usepackage[defaultcolor=red]{changes}

\usepackage{newtxtext,newtxmath}
\usepackage{comment}

\newcommand{\noopsort}[1]{}

\def\al#1\eal{\begin{align}#1\end{align}}
\def\beq#1\eeq{\begin{equation}#1\end{equation}}
\def\wperp{\tilde{w}_\perp}

\def\rmin{r_\mathrm{min}}
\def\toff{t_\mathrm{off}}
\def\tmin{t_\mathrm{min}}

\newcommand{\RNum}[1]{\uppercase\expandafter{\romannumeral #1\relax}}

\title[Advancing Stellar Streams as a Dark Matter Probe --- \RNum{1}:  Effects of Subhalo Density Profile]{Advancing Stellar Streams as a Dark Matter Probe --- \RNum{1}:  Effects of Subhalo Density Profile}

\author[Menker \& Benson]{
Paul Menker,$^{1,2}$\thanks{E-mail: pmenker@usc.edu}
and Andrew Benson$^{2}$
\\
$^{1}$University of Southern California, 3551 Trousdale Pkwy, Los Angeles, CA 90089, USA\\
$^{2}$Observatories, Carnegie Institution for Science, 813 Santa Barbara Street, Pasadena, CA 91101, USA
}

\date{Accepted XXX. Received YYY; in original form ZZZ}

\pubyear{2023}

\begin{document}
\label{firstpage}
\pagerange{\pageref{firstpage}--\pageref{lastpage}}
\maketitle

\begin{abstract}
Stellar streams, long, thin streams of stars, have been used as sensitive probes of dark matter substructure for over two decades. Gravitational interactions between dark matter substructures and streams lead to the formation of low-density ``gaps'' in streams, with any given stream typically containing no more than a few such gaps. Prior models for the statistics of such gaps have relied on several simplifying assumptions for the properties of the subhalo population in the cold dark matter scenario. With the expected forthcoming increase in the number of streams and gaps observed, this work develops a more detailed model for the statistics of subhalos interacting with streams and tests some of the assumptions made in prior works. Instead of using simple fits to N-body estimates of subhalo population statistics at $z=0$ as in previous work, we make use of realizations of time-dependent subhalo populations generated from an entirely physical model, incorporating structure formation and subhalo orbital evolution, including tidal heating and stripping physics, which has been carefully calibrated to match results of cosmological N-body simulations. We find that this model predicts 20\% more gaps  (up to 60\% for deep gaps) on average in Pal-5-like streams than prior works.
\end{abstract}

\begin{keywords}
cosmology: theory, dark matter
\end{keywords}

\section{Introduction}\label{sec:introduction}

The nature of dark matter\footnote{See \cite{annika1} for an accessible introduction}, first postulated in the 1930's \citep{zwicky1, zwicky2}, and widely accepted in the 1970s \citep{rubin1, rubin2}, and which makes up 24\% of the universe's energy budget, remains an unsolved problem after over 90 years. ``Baryonic'' candidates have effectively been ruled out,\footnote{Aside from small remaining mass windows for primordial black holes, as explained in \cite{annika2}, and the references therein.} and therefore new particles must be added as extensions to the Standard Model of particle physics \citep{weinberg1}.  While alternatives to dark matter, such as modified gravity, have been put forward \citep{mond1, mond2}, in this work, we assume that dark matter is a particle.

Given the ``particle zoo'' of proposed dark matter candidates \citep[][and references therein]{annika2}, which span at least $30$ orders of magnitude in particle mass and $50$ orders of magnitude in interaction strength, it is natural to perform Earth-based experiments for specific candidates. Each model has characteristic dark--light interactions, which are, in principle, testable. ``Missing energy'' experiments in particle colliders have searched for indirect signals of a `Lightest Supersymmetric Particle' \citep{string3, susy2}, however, no anomalous signal has been found \citep{string4, susy3}.  Noble gas scintillators have ruled out the ``WIMP Miracle'' \citep{wimp1, xenon1, xenon2, xenon3}, and anomalous Solar neutrino signals have not survived reanalysis \citep{xenon4, xenon5, xenon6, xenon7}. While new interactions are always possible at higher energy scales \citep{extended}, the LHC is already halfway through its third run, \citep{cern1} and no new signals have been found. It is perfectly allowable for \emph{no} dark-standard model interactions to occur \citep{scalar}, and so it remains imperative to consider astrophysical tests. 

Given that dark matter was first observed in large-scale structures, it is natural that astrophysics has been a fruitful source of tests. MACHOS have been ruled out \citep{annika1, annika2},  primarily through gravitational lensing of the Magellenic Clouds \citep{MACHOLensing1, MACHOLensing2} and abundances of wide binaries \citep{MACHOBinaries}.  Cold Dark Matter (CDM; \citealt{cdm1}) has triumphed over Hot Dark Matter \citep{zeldovich1} on large scales, and a ``$\Lambda$CDM paradigm'' has been established \citep{dodelson}.  Today, CDM is considered consistent with all observations on mass scales $\gtrsim 10^8 \mathrm{M}_\odot$ (\citealt{nadler1}, \citealt{annika2}).\footnote{Note that discrepancies, such as the core-cusp problem, have challenges CDM, but these issues have largely been resolved \citep{core-cusp1, core-cusp2, core-cusp3}.}  However, other models, such as Warm Dark Matter(WDM) and Self-Interacting Dark Matter (SIDM), also match current constraints.  CDM predicts the existence of overdense dark matter subhalos down to very small mass
scales \citep{reviewerCDM1, reviewerCDM2, 2008MNRAS.391.1685S,subhalosExist,2024MNRAS.528.7300Z}, while, for example, WDM models exhibit a cut off in the mass spectrum of subhalos. Therefore, the key to our understanding of dark matter is the low-mass subhalo population, where the degeneracy between models is strongly broken. 

Stellar streams may provide new constraints on dark matter properties on small mass scales.  They have been studied in detail for more than two decades and yet have not been widely appreciated as potentially competitive constraints on dark matter.  Formed through tidal stripping of dwarf galaxies and globular clusters (GCs), phase-space constraints transform this stripped material into long, thin streams of stars \citep{early1, early2, early3, early4, early5}.  An essential feature is the uniform motion, or ``coldness'' of these streams, with velocity dispersions as low as $\sim$1--2~km/s in globular cluster streams, and closer to $\sim$10--20~km/s in dwarf galaxy streams. Therefore, these streams can act as a blank canvas for gravitational interactions, recording details of all perturbations since the stream's formation.  In this work, we aim to build upon the large body of pre-existing literature, such as \cite{firstGaps1, firstGaps2, kupper1, kupper2, carlberg3, carlberg1, carlberg2020, fardal, gaps5, erkal1, erkal2, gaps3, erkal3, gaps11, gaps1, gaps2, spur, gaps9, gaps8, gaps10, gaps7, gaps6}, and in particular \cite{erkal3}, to provide a more realistic understanding of gap formation through semi-analytic methods, that can be applied at scale. 

Given that future surveys, such as those to be conducted by the Nancy Grace Roman Space Telescope \citep{future1,future2}, and the Via Project \footnote{\href{https://via-project.org/}{\tt https://via-project.org/}}, will take the current stream catalog from $\sim 250$ streams \citep{recent1, streamfinder1, streamfinder2, cats}, to several thousand, and provide unprecedented resolution currently available only for streams such as  GD-1\citep{spur}, it is imperative that our theoretical models can fully capitalize on this data.  Different approaches have been taken, but here we employ a semi-analytic modelling of gaps, which has several advantages.  Most importantly, semi-analytic calculations are faster than N-body simulations and more accurate than analytical models. Given how few gaps exist per stream---meaning that a single stream will provide only a very low signal-to-noise constraint on dark matter physics, it is crucial to generate many ($\sim 10^6$) realizations of the Milky Way and its subhalo population to achieve results that are accurate at the $\sim 1\%$ level.  Additionally, the equations derived from this approach can provide model-dependent insights, allowing stream data to be maximally utilized.

This work is the first of a series of projects in which we intend to further semi-analytical modelling of gaps and ultimately hope to make realistic constraints on dark matter averaged over the entire ensemble of streams. To this end, this work makes improvements in a number of ways (including a semi-analytic subhalo population, realistic mass profiles, and several improvements to gap modeling that cause a $\sim250-3000$ times speed-up of post-processing). This paper is focused on establishing the basic model rather than providing final results and employs a number of approximations, each of which we intend to assess in detail in subsequent works.  The methods employed are explained in \S\ref{sec:methods}. Gap results for the Pal-5 stream are computed in \S\ref{sec:results}, and are compared to previous work \citep{erkal3}, and briefly generalized in \S\ref{sec:discussion}. These results are summarized, along with an outline of future improvements, in \S\ref{sec:conclusions}.

\section{Methods}\label{sec:methods}
\subsection{Overview}\label{sec:overview}
The methods presented in this paper can be neatly divided into two parts. First, in \S\ref{sec:galacticus}, subhalo populations are generated semi-analytically using {\sc Galacticus}. Then, in \S\ref{sec:preparations}--\ref{sec:gaps}, gap formation is modeled as a scattering event.

To the authors' knowledge, this is the first time\footnote{After submission of this work, \cite{Adams2024} also described a semi-analytic approach to generating subhalo-stream interaction statistics.} such an approach has been employed for stellar streams, balancing between the earlier analytic work of \cite{yoon, carlberg3, erkal1, erkal2, erkal3} and later, fully numerical work such as \cite{fardal,  streamsPowerSpectrum, improvement2, gaps10, reviewerStreams1}.  As we will see, extremely large sample sizes must be generated to achieve the statistical power necessary to separate dark matter candidates. This work alone uses $1.6 \times 10^6$ realizations of the Milky Way, which reduces statistical uncertainty to $\sim 1\%$. While these simplified models of gap formation will never match direct particle-spray simulations, it is also unlikely that those detailed models can be run at the necessary scale to constrain non-CDM candidates beyond current bounds.  If analytic approaches lack sufficient accuracy, and full N-body simulations may lack the requisite speed, the motivating hope behind a semi-analytic approach is that it can work ``well enough, quickly enough.''  While this assumption has yet to be fully verified, and a truly accurate semi-analytic approach requires improvements beyond that of a single paper, this regime has certainly been underutilized and is at least worthy of investigation.

Once the subhalo population has been generated, interactions are modeled as impulsive scattering events.  The limits of these assumptions are discussed. This methodology is built upon the work of \cite{carlberg3}, \cite{erkal1}, and \cite{erkal3}, with several substantial improvements in both accuracy and speed and new, more tractable forms of the velocity kicks. After converting {\sc Galacticus}'s output into stream-specific variables in \S\ref{sec:preparations}, gaps are computed in \S\ref{sec:gaps}. Interactions are mediated through velocity shifts in the stream in \S\ref{sec:kicks}, modified orbits are calculated in \S\ref{sec:orbits}, and under-dense regions, known henceforth as ``gaps,'' are modeled in \S\ref{sec:gapFormation}. In this work, we consider only subhalos with impact-time bound masses between $10^5 \mathrm{M}_\odot$ and $10^9 \mathrm{M}_\odot$, which was previously believed to be the gap-forming region of the subhalo population.\footnote{Smaller subhalos were believed to be weak perturbers unable to interact significantly, and larger subhalos were too infrequent to matter.}  However, as discussed in \S\ref{sec:discussion}, the realistic, cuspy profiles employed here suggest that smaller subhalos should be pursued in future work. Given the desire for tests of the subhalo mass function in this range, this is an extremely exciting development.

The mass profile of a subhalo, how its matter is distributed within its bound radius, can vary both between candidates and with approximations employed in the simulation. CDM strictly follows a ``tidally stripped NFW profile'', \citep{tidallystripped1, tidallystripped2, 2022MNRAS.517.1398B} which is employed here for the first time in gap-formation work. During halo assembly, accreted halos oftentimes have their outer density profile strongly perturbed by the tidal field of their host halo, resulting in ``tidally stripped'' subhalos. Additionally, gravitational tidal shocks can heat the inner regions of the subhalo, leading to further changes in the density profile \citep{gnedin_effects_1999}.\footnote{We note that there has been significant discussion in the recent literature on the question of whether tidal heating is actually relevant in subhalo cusps. For example, \citeauthor{2023MNRAS.521.4432S}~(\citeyear{2023MNRAS.521.4432S}; see also \citealt{2021arXiv211101148A,2024ApJ...971...91R}) describe a model that matches the tidal tracks of CDM subhalos through a pure tidal stripping (i.e. no tidal heating) approach, although they acknowledge that the tidal heating approach used here \citep{2022MNRAS.517.1398B} may be consistent with their results (the two methods simply approaching the problem from different directions). In any case, the key point for the present work is that the tidal heating model results in good agreement with high-resolution N-body simulations of CDM subhalo density profiles \citep{2022MNRAS.517.1398B}.}

Alongside this profile, two other realistic profiles and three toy models are employed. Each is either included for historical reasons or to shed some light on the model-dependence of gap formation.\footnote{Understanding this is a major goal of this program. It makes no difference how realistically we can simulate and observe gap formation if these gaps do not vary between models.} The physically plausible models are a Plummer sphere,\citep{Plummer1, Plummer2} which has been the standard in the field\citep{erkal1, erkal3, Adams2024} and a pure NFW profile \citep{NFW}, which is a more realistic analytical approximation of a CDM subhalo.

Additionally, a point mass is modeled, which was used in early modeling \citep{yoon, carlberg3}, and represents the upper bound on the interaction strength of mass profiles. A thin shell, where all the mass is concentrated directly at the subhalo radius, is used to provide a lower bound on the gaps a subhalo with fixed mass and radius can create. Finally, a constant density profile represents a middle ground between these two, lacking the cuspy center of NFW-like profiles.

\subsection{Generating Subhalo Populations}\label{sec:galacticus}

Every gravitationally accurate realization of a galaxy must include its dark matter halo, and all the substructure contained within. Because dark matter is not directly observable, and so can not be fully mapped within our galaxy,\footnote{Of course, larger dark matter substructures---such as the subhalo hosting the LMC---can be measured with reasonable precision, such that their effects can be accounted for directly \protect\citep{Shipp_2021}. We will comment further on this in \S\ref{sec:discussion}.} one must model its presence statistically to make predictions about gaps. To make accurate and statistically-representative predictions we generate random realizations of subhalo populations using a physics-based approach. Specifically, we make use of the {\sc Galacticus} semi-analytic model \citep{galacticus} to predict the assembly and orbital evolution (and eventual destruction) of populations of subhalos without the computational cost of full N-body modelling. For a full description of how {\sc Galacticus} models subhalo physics we refer the reader to \cite{2014ApJ...792...24P}, \cite{2020MNRAS.498.3902Y} and \cite{2022MNRAS.517.1398B}.

Briefly, {\sc Galacticus} starts from a Milky Way-mass ($10^{12}\mathrm{M}_\odot$), halo at $z=0$, and constructs the assembly history of that halo by building a merger tree using the algorithm of \cite{2008MNRAS.383..557P}, resolving progenitor halos down to a mass of $10^5\mathrm{M}_\odot$. For computational efficiency, we do not follow every branch of the tree, but instead subsample branches. This allows us to avoid having to track the evolution of huge numbers of low-mass halos, while having only a small number of higher-mass halos. Specifically, at each branching of the merger tree, we randomly decide whether or not to follow the lower mass branch with probability
\begin{equation}
    p(M) = \left\{ \begin{array}{ll} (M/10^9\mathrm{M}_\odot) & \hbox{if } M < 10^9\mathrm{M}_\odot, \\ 1 & \hbox{if } M \ge 10^9\mathrm{M}_\odot, \end{array} \right.
    \label{eq:subsample}
\end{equation}
where $M$ is the mass of the lower mass branch. Of course, this subsampling affects the statistics of subhalo-stream interactions which we wish to calculate in this work. To correct for this we evaluate a weight, $w_\mathrm{i}$, for each halo/subhalo which accounts for this subsampling. When a branching occurs in the merger tree, and the branch is retained, we assign the halo in that branch a weight $w_\mathrm{i} = w_\mathrm{j}/p(M_\mathrm{i})$ where $w_\mathrm{j}$ is the subsampling weight of the parent halo, $M_\mathrm{i}$ is the mass of the new branch, and $w_1 = 1$ (the weight of the root halo in the tree). We have verified that, when applying these subsampling weights, this approach results in a distribution of subhalo masses (and other properties) that precisely matches the original, non-subsampled algorithm.

Each halo in the resulting merger tree is assumed to be described by an NFW density profile\footnote{Using an \cite{1965TrAlm...5...87E} density profile would provide a slightly more accurate description of the distribution of mass in halos \citep{2004MNRAS.349.1039N,2008MNRAS.387..536G,2008MNRAS.388....2H}. However, the current {\sc Galacticus} model is calibrated assuming NFW halos, so we retain that assumption here.} \citep{NFW}, with a virial density contrast given by the spherical collapse model (as appropriate to the dark energy cosmological model used in this work; \citealt{1996MNRAS.282..263E,2005A&A...443..819P}), and a concentration determined using the model of \cite{2021ApJ...908...33J} which computes the NFW scale radius from each halo's assembly history.\footnote{We find these Milky Way halos to have a median concentration of $c=8.33$ at $z=0$ when using this model. This concentration calculation could, in principle, be affected by our choice to subsample branches of the merger tree, as it uses the properties of merging halos to compute the evolution of concentration.  To avoid this problem, we initially retain the ``stub'' of each branch (i.e. the halo of that branch which is about to merge with its parent halo, but none of the progenitors of that halo) that is rejected by our subsampling procedure. These ``stub'' branches contain the information needed to apply the concentration calculation of \cite{2021ApJ...908...33J}, ensuring that the resulting concentrations are unaffected by subsampling. Once concentrations have been computed, the stub branches are finally removed from the merger tree. With this approach, we find that the actual effect of subsampling on concentrations of halos is minor. For example, we find that the mean concentration of the Milky Way halo is shifted by less than 0.08~dex when we include subsampling, while there is almost no change in the halo-to-halo variance in the concentration.}  Halo spins are determined from the assembly history of each halo using the model of \cite{2020MNRAS.496.3371B}.

The halos in each merger tree are then evolved forward in time. When two halos in the tree merge, the lower mass halo becomes a subhalo in the higher mass halo. As the tree continues to evolve, more and more halos merge, resulting in a population of subhalos (and sub-subhalos, etc.). {\sc Galacticus} follows the orbital evolution of each subhalo in the time-evolving potential of its host halo. Initial orbital parameters are assigned from the cosmological distribution measured by \citeauthor{2015MNRAS.448.1674J}~(\citeyear{2015MNRAS.448.1674J}; with best-fit parameter values as determined by \cite{2020MNRAS.496.3371B} and including the correlation with the host halo spin as found by those authors) at the time of merging. The orbital motion is then integrated forward in time,e including the effects of dynamical friction, tidal stripping, and tidal heating \citep{2014ApJ...792...24P}.  The latter modifies the original density profile of a subhalo according to a simple heating model \citep{2022MNRAS.517.1398B}. 

If a halo containing subhalos merges into a yet larger halo, thereby becoming a subhalo itself, the original subhalos are now sub-subhalos.  Such a sub-subhalo feels the gravitational acceleration (and tidal forces) only from its direct subhalo host---the gravitational accelerations from any higher level hosts are ignored. The orbit of the sub-subhalo is, of course, computed relative to the center of the subhalo, which in turn continues to orbit around its own host halo. These sub-subhalos remain bound to their original host halo (and continue to orbit within it) until and unless they stray outside the instantaneous tidal radius of that host. When this happens, the sub-subhalo (which we will refer to as halo ``A'') is assumed to be tidally stripped from its subhalo host (halo ``B''), and becomes a subhalo in the host halo (halo ``C'') of halo B, with an initial position and velocity equal to the vector sum of the orbital position/velocity of A in B, plus the orbital position/velocity of the B in C. While this is, of course, an approximation (the sub-subhalo should feel gravitational accelerations from all levels of host (sub)halo at all times), this is a computationally-efficient approach. A more detailed model, accounting for some contributions from higher-level hosts, is being developed (Du et al., in preparation). At any given time, {\sc Galacticus} thereby provides full details (current bound masses, orbital positions and velocities, and density profiles) of all subhalos orbiting within the Milky Way halo. This model has been calibrated to accurately match subhalo mass functions and radial distributions measured from cosmological N-body simulations \citep[][see also \citealt{2023ApJ...945..159N}]{2020MNRAS.498.3902Y}, and to match the evolution of tidally-heated halos in idealized simulations \citep{2022MNRAS.517.1398B}. As such, it provides an accurate and detailed description of cosmological subhalo populations.

In particular, this model accurately matches the tidal tracks and density profile evolution of subhalos measured in high-resolution N-body simulations \citep{2024PhRvD.110b3019D}, ensuring that the internal structure of the model subhalos (which will be important in determining the velocity kicks that a subhalo imparts to stream particles) is realistic. As found by several authors \citep{10.1093/mnras/stab1215,2022MNRAS.517.1398B, reviewerGal1}, cuspy profiles such as NFW are never fully disrupted by tidal forces---this is true in the model implemented in {\sc Galacticus}. For practical purposes, however, we stop tracking subhalos for which the bound mass falls below $10^5\mathrm{M}_\odot$.  As we will show in Figure~\ref{fig:countDecade} this has only a minor effect on the overall number of stream gaps produced. 

For this work, we perform dark matter-only simulations---all baryonic physics is switched off in {\sc Galacticus} such that no galaxies are formed. This allows for comparison with earlier works which were based on fits to dark matter-only N-body simulations. This also means that the host potential (in which the subhalos orbit) has no contribution from the Milky Way galactic disk or bulge. The presence of the galactic potential would enhance tidal stripping and heating of subhalos, and would plausibly lead to a reduction in the number of subhalos in the inner region of the Milky Way halo (the region relevant for subhalo-stream interactions in this work) of 30\% \citep{2022MNRAS.509.2624G}. Of course, {\sc Galacticus} can incorporate a detailed treatment of galaxy formation physics, including the effects of the Milky Way galactic potential on subhalo population statistics, which may have important implications for subhalo populations and the statistics of gaps in stellar streams (see \S\ref{sec:discussion} for further discussion of this point)---this will be the focus of a future work.

To model the time evolution of the subhalo population, {\sc Galacticus} models subhalos as tidally stripped NFW density profiles. For the other density profiles that we consider (Plummer, NFW, and the three toy models) the profiles are constructed from the subhalo masses tracked by {\sc Galacticus} \emph{including} all tidal effects. Thus, changing the density profile does not change the distributions of subhalo masses or orbital parameters. This is intentional and allows us to isolate the effects of changing the density profile alone.

For this work, we focus on the Pal-5 stream \citep{odenkirchen, pal5}. We generate 4,000 merger tree realizations of our Milky Way galaxy's halo, taking snapshots of the subhalo population at 1,000 timesteps, with timesteps evenly spaced between the stream's formation and the present. Pal-5 is 3.4~Gyr old \citep{pal5age}, 9~kpc long \citep{pal5length}, and at an average distance of 13~kpc \citep{pal5r} from the Galactic centre, with length growing approximately linearly in time.\footnote{See, for example, Figure~2 of \protect\cite{streamsPowerSpectrum}.  We ignore the oscillations in stream length due to the eccentricity of the stream orbit as, in this work, we approximate the stream with a circular orbit.} Because the timestep $\Delta t,$ between simulations is of order 1~Myr, which is much less than the dynamical time in the halo, our subhalo population is well-resolved in time.

Each timestep contains exhaustive information about the subhalos present, in particular their existence, mass, orbital position and velocity, subsampling weight (as described above), concentration, and virial radius, which are used for our NFW model. Additionally, for each subhalo, a tabulated density profile is output, which describes the tidally stripped NFW profile. From this data, several calculations are performed. First, the population is cleaned of any halos that are not subhalos of the main branch of the merger tree.\footnote{At times $z>0$, the outputs from our simulations also include halos and subhalos that have not yet merged into the Milky Way progenitor halo. These are excluded as they, by definition, can not have a close encounter with the stream in the Milky Way progenitor halo at this time.} Then, because we assume that the subhalo population is not correlated with the orientation of the stream, and model the stream with a circular orbit, we can freely rotate the system.\footnote{This will no longer be true when baryons are included in the simulation in future works, which break the spherical symmetry of our host halo.} This effectively increases the sample size of subhalos that interact strongly with the stream and is performed for 400 isotropically random rotations.  Additionally, taking further advantage of this isotropy of the subhalo system in the simulated mass range, the stream is extended from an arc to a full circle for the purposes of finding close encounters with subhalos (the resulting gap statistics will be weighted accordingly to the actual length of the stream at each snapshot, by multiplying the contribution of each gap by a factor of $l(t)/2 \pi r_0$, where $l(t)$ is the length of the stream as a function of time, and $r_0$ is the orbital radius of the stream). Note that all models use the same subhalo population and differ only in the density profiles of their subhalos.

 \subsection{Preparing Subhalos}\label{sec:preparations}
A chief concern of this approach is speed. Appendix \ref{sec:impactAppendix} describes a series of filters to remove unwanted subhalos, speeding up this stage of the post-processing from $\sim 1$ day per timestep to $\sim 3$ minutes per timestep. This section will assume all filters have already been applied. Discussions of these improvements are primarily discussed in Appendices \ref{sec:impactAppendix}-\ref{sec:hypergeometricAppendix}, however Table \ref{table:speedups} provides a brief summary of these improvements.
\begin{table*}
\begin{center}
\begin{tabular}{c c c c c c c}
    \hline
    Section  & Main improvement & 
    Approximate speedup & \\
    \hline
    Appendix A    &   Filters based on maximal interaction strength & 10$\times$  \\ 
    Appendix B     &     Exploits symmetries in velocity kicks      & 6$\times$  \\
    Appendix C      &    Solve velocity kicks analytically & 120$\times$ (when applicable)  \\
    Appendix D       &     Evaluate velocity kicks using hypergeometric functions & 16 $\times$ (for non-analytic models) \\
    \hline
\end{tabular}
\caption{Summary of improvement and average speedup resulting from each of the optimizations made in this work (as described in the associated appendix listed in the first column).}
\label{table:speedups}
\end{center}
\end{table*}

Due to the symmetries discussed in \S\ref{sec:galacticus}, the stream will be treated as a circle in the $x$--$y$ plane. The orientation aligns with convention, providing ease of notation when engaging with the literature.  Extending the stream from a circular arc to a full circle, however, is new and allows us to obtain more statistical power from each realization of the subhalo population. To compensate, the subhalo weights, first mentioned in \S\ref{sec:galacticus}, are updated as 
\begin{equation}
    w \rightarrow w\times \frac{ l_\mathrm{stream}}{2\pi r_\mathrm{stream}}\times\frac{t_\mathrm{impact}}{T_\mathrm{stream}},\label{eq:weights}
\end{equation}
where $t_\mathrm{stream}$ and $T_\mathrm{stream}$ are respectively the time since the subhalo's impact with the stream and the stream's age at $z=0$. That is, we assume uniform growth of the stream and that the total number of encounters scales with the stream's length at a given time. Then, each of the 4,000 realizations from {\sc galacticus} is randomly rotated 400 separate times about the $x$ and $y$ axes. These rotated samples can be treated as approximately independent, increasing our sample size to 1,600,000 merger trees. Filters in \S\ref{sec:impactAppendix} are performed at various stages to manage file sizes and computational speed. 

We treat stream--subhalo interactions as a scattering process, governed by an impact parameter: $b$. At any given time, the distance between a point along the stream, $\mathbf{r'}(\phi)$, and the subhalo at $\mathbf{r}(t)$ is given by 
 
\begin{equation}
    \mathbf{d}(\phi, t) = \mathbf{r'}(\phi) -\mathbf{r}(t).
\end{equation}
Following \cite{erkal1}, we assume an impulse approximation, specifically that subhalos are on straight line trajectories, and the interactions happen faster than the reaction time of the stream, and can be modeled as instantaneous encounters during which the positions of stars in the stream do not change. Using the first of these assumptions, we can expand
\begin{equation}
    \mathbf{d}(\phi, t) = \mathbf{r'}(\phi) -\left(\mathbf{r_0} + \mathbf{v_\mathrm{0}} t \right), \label{eq:distance1}
\end{equation}
where $r_0$ and $v_0$ denote the initial position and velocity of the subhalo at the current timestep, and t is defined relative to the start of that timestep.\footnote{Note that technically, eqn~(\ref{eq:distance1}) neglects the rotation of the stream, but given the symmetry of the unperturbed stream, this is unimportant. Once the impact position is fixed, we will use relative velocity to model the impact itself.} The subscripts will be dropped in subsequent equations for simplicity. 

For any point along the stream, the minimum distance to the subhalo will occur when 
\begin{equation}
\frac{d |d(\phi, t)|}{dt} =0,
\end{equation}
namely at 
\begin{equation}
      t_\mathrm{min}(\phi) = \frac{\left(\mathbf{r'}(\phi) -\mathbf{r}\right)\cdot \hat{\mathbf{v}}}{v}.  \label{eq:impactTime}
\end{equation}

However, we are not interested in the distance to any point along the stream. Instead, we are interested in the smallest distance between any point along the stream and the subhalo, at any time: 
\begin{equation}
b \equiv \mathrm{min}[d(\phi, t_\mathrm{min}(\phi)].
\end{equation}
Substituting eqn.~(\ref{eq:impactTime}) into eqn.~(\ref{eq:distance1}), we find
\begin{equation}
    \mathbf{d}_\mathrm{min}(\phi) = \Delta \mathbf{r}(\phi) -\left( \Delta \mathbf{r}(\phi)  \cdot \hat{\mathbf{v}}\right) \hat{\mathbf{v}},  \label{eq:distance2}
\end{equation}
where
$\mathbf{\Delta r}(\phi) = \mathbf{r'}(\phi) -\mathbf{r}$, and $\hat{\mathbf{v}} = \frac{\mathbf{v}}{v}$.\footnote{Effectively, displacement along the trajectory of the stream can be removed through its motion, but displacement perpendicular to it cannot.}

Next, we find the norm:
\begin{equation}
    d_\mathrm{min}(\phi) =  \sqrt{\Delta r^2(\phi) - \left(\Delta \mathbf{r}(\phi)\cdot \hat{\mathbf{v}}\right)^2},\label{eq:distanceFinal}
\end{equation}
where the minus sign arises from cancellations due to cross terms.  Formally, eqn.~(\ref{eq:distanceFinal}) can be minimized directly to find $b=\mathrm{min}(d_\mathrm{min}(\phi))$. However, global minima solvers are famously both slow and inaccurate. Instead, Appendix \ref{sec:impactAppendix} describes more filters that are applied and a procedure to solve this equation using quartic roots of Weierstrass functions.

Regardless of the method, once $b$ is found, other relevant parameters can be computed straightforwardly. In line with previous work \citep{yoon, carlberg1, erkal1}, the $x$--$y$ plane is now rotated by $\frac{\pi}{2}-\phi_\mathrm{min}$ so that the subhalo is now in the $x$--$z$ plane at the time of closest approach. Taking the stream, then, to be traveling in the $+y$ direction, we can compute 
\begin{equation}
\begin{split}
    w_\perp = \sqrt{v_x^2 + v_z^2},\\
    w_\|  = v_y - v_\mathrm{str},\\  
    w = \sqrt{w_\perp^2 + w_\|^2}   \label{eq:w},
\end{split}
\end{equation}
where $\mathbf{v}$ is the velocity of the subhalo, and $v_\mathrm{str}$ is the velocity of the stream.\footnote{Unlike in Appendix~\ref{sec:impactAppendix}, we are no longer working in cylindrical coordinates. Therefore, note that generically $v_y\neq 0$, and $v_x$ can be negative.}  For Pal-5, we assume a potential resulting in a circular velocity of $v_\mathrm{str} =220$ km/s at the location of Pal-5, in line with \cite{erkal3}. 

An angle, $\alpha$, can be used to split the perpendicular coordinates into $x$ and $z$ components, with 
\begin{equation}
\begin{split}
(b_x, b_z) = \left( b\cos\alpha, b\sin\alpha \right),\\
(w_x, w_z) = \left( -w_\perp \sin\alpha, w_\perp\cos\alpha \right).
\end{split}
\end{equation}

\subsection{Calculating Gap Sizes}\label{sec:gaps}

With a proper coordinate system established, we will now model the stream--subhalo interaction.  This approach is in the spirit of \cite{erkal1} and subsequent papers, however several improvements have been made.  Specifically, \S\ref{sec:kicks} represents velocity kicks in a new way,\footnote{Note that this new form is mathematically equivalent, but several times quicker to use computationally, and is far easier to work with symbolically, as discussed in Appendices \ref{sec:velocityKickAppendix}-\ref{sec:modelAppendix}.} \S\ref{sec:orbits} includes an exact orbital shift,\footnote{As discussed below, this shift still employs first order approximations, but eqn.~(\ref{eq:integral}) is solved exactly, unlike previous work.} and \S\ref{sec:gapFormation} presents gap sizes valid at all subsequent times.\footnote{\cite{erkal1} presented separate formulae for early, intermediate, and late times in the evolution of the gaps. By evaluating gap sizes numerically, these approximations are not necessary, and our results are valid during transitions between these regimes.}

Turning towards the perturbation process as a whole, it can effectively be broken into three steps:
\begin{enumerate}
    \item (\S\ref{sec:kicks}) a scattering interaction occurs, resulting in position-dependent ``velocity kicks'' in the stream;
    \item (\S\ref{sec:orbits}) these velocity kicks change the orbit of each point near the perturber, creating a transfer function of stream material away from the impact centre;
    \item (\S\ref{sec:gapFormation}) this transfer function can be inverted to compute the resulting density of the stream, and thus the size and number of ``gaps'' that form.
\end{enumerate}

\subsubsection{Velocity Kicks}\label{sec:kicks}

The interactions between a subhalo and a stream are strictly gravitational. This work assumes that interactions take place over a very short period of time (relative to the orbital timescale of stream particles) and that the stream does not have time to adjust during the interaction. The total interaction can be written as:\footnote{We note that we extend the limits of the integral in eqn.~(\ref{eq:velocityKickBasic}) to $\pm\infty$. Of course, the expression for $r(y,t)$, derived under the assumption of constant velocities will be highly inaccurate in this limit. Some authors \cite[e.g.][]{Adams2024} instead truncate the integral beyond some distance. Given the $1/r^2$ nature of gravitational forces, we find that truncating the integral or not results in sub-percent differences in the velocity kick.} 

\begin{equation}
\mathbf{\Delta v}= \int_{-\infty}^{\infty} \frac{-\mathrm{G} M_\mathrm{enc}(r)}{r^2(t)} \hat{\mathbf{r}}(t)\, \mathrm{d}t,
    \label{eq:velocityKickBasic}    
\end{equation}
with 
\begin{equation}
r(y,t) \equiv |r_\mathrm{sub}(t) - r_\mathrm{stream-point}(y,t)|.\label{eq:rBasic}
\end{equation}
By design, the displacement between the subhalo's centre of mass, and a point along the stream, $y$ is minimized at $y=0$ and $t=0$.

Now, as in \S\ref{sec:preparations}, we assume that the perturber has a linear trajectory and that the stream can locally be regarded as straight near the point of impact. While these assumptions will be relaxed in future work, they allow for a tractable and relatively accurate framework at present. Using these assumptions, we find
\begin{equation}
\mathbf{\Delta v}= \int_{-\infty}^{\infty} \frac{-\mathrm{G} M(r)}{(y + w_\parallel t)^2 + w_\perp ^2 t^2 +b^2} \hat{\mathbf{r}}\, \mathrm{d}t, 
    \label{eq:velocityKickTime}    
\end{equation}
or by using
\begin{equation}
    \hat{\mathbf{r}} = \left(\frac{b_x + w_x t}{r},\frac{y + w_\| t}{r}, \frac{b_z + w_z t}{r} \right), \label{eq:componentsEarly}
\end{equation}
one can rewrite each component of the velocity kick as 
\begin{equation}
\Delta v_i= \int_{-\infty}^{\infty} \frac{-\mathrm{G} M(r) \left( c_1 + c_2 t\right)}{\left((y + w_\parallel t)^2 + w_\perp ^2 t^2 +b^2\right)^{3/2}}  \mathrm{d}t, \label{eq:kickStartingPoint}
\end{equation} 
where $c_1, c_2$ are geometric factors that vary for each component.
Exploiting symmetries, details of which we provide in Appendix~\ref{sec:velocityKickAppendix}, these velocity kicks can be expressed in the form 
\begin{align}
    \Delta v_i(y) &=&-\frac{2\mathrm{G}M_\mathrm{tot}X_i(y)}{w \rmin^2(y)} \times\nonumber\\
    &&\left(1  - \frac{\sqrt{\tilde{R}_\mathrm{max}^2- 1}}{\tilde{R}_\mathrm{max}} + 
    \int_{1}^{\tilde{R}_\mathrm{max}} \frac{M(\tilde{r})}{ \tilde{r}^2 \sqrt{\tilde{r}^2- 1}}  \mathrm{d}\tilde{r} \right) \label{eq:integralFinal}, 
\end{align}
with 
\begin{equation}
\begin{split}
X(y) &\equiv& b_x + w_x\toff(y),\\
Y(y) &\equiv& y \wperp^2,\\
Z(y) &\equiv& b_z + w_z\toff(y), \label{eq:componentsFinal}
\end{split}
\end{equation}
$\tilde{r} = \frac{r}{\rmin}$, $\tilde{M}(\tilde{r}) = \frac{M(\tilde{r}\rmin)}{M_\mathrm{tot}}$, $\tilde{R}_\mathrm{max} = \mathrm{min}\left(R_\mathrm{halo}/\rmin, 1\right),$ $\rmin(y) = \sqrt{b^2 + y^2 \wperp^2}$, $\toff = \frac{-y\tilde{w_\parallel}}{w}$, and $\tilde{w}_{\parallel, \perp} = \frac{w_{\parallel, \perp}}{w}$. A derivation and more detailed explanation can be found in Appendix \ref{sec:velocityKickAppendix}. However, the crucial points are that all velocity components only require a single integral to be computed, that these integrals are symmetric in $y$, and that they only need to be performed for impacts deeper than the perturber's radius.\footnote{This is only true for bound subhalo mass profiles. Note that while NFW-like profiles formally have no upper radius, in practice they are truncated at the virial radius to avoid an unbounded total mass. Therefore, this approach is employed for every mass profile aside from a Plummer Model.}

Previous work \citep{erkal2} has found a degeneracy in the velocity kicks for the Point and Plummer models (eqns.~\ref{eq:point}--\ref{eq:Plummer}). Specifically, a more massive, high-speed perturber could result in the same kick as a less massive, lower-speed perturber, leaving velocity kicks unchanged if $[M_\mathrm{tot}, w] \rightarrow [\lambda M_\mathrm{tot}, \lambda w]$. In detail, however, this degeneracy will be broken for a Plummer model due to the dependence of the Plummer radius on mass: $r_\mathrm{p}\propto M_\mathrm{tot}^{0.5}$ \citep{erkal3}. \cite{erkal2} do not account for this dependency in their equations 15--17. Of course, this requires the adoption of some model for the $r_\mathrm{p}(M_\mathrm{tot})$ relation, but, in principle, one could uniquely reconstruct a perturber's total mass, velocity, and mass profile from a gap itself, \footnote{In \S \ref{sec:orbits}, we will see that no observable depends on $w_z$ to first order, so there is a trivial degeneracy here for all candidates.} for any model except a point mass (which exhibits scale-free behavior), including the other mass profiles in this paper (eqns.~\ref{eq:IShell}, \ref{eq:IConstant}, \ref{eq:NFWKick} and Appendix \ref{sec:hypergeometricAppendix}). However, in practice, observational limits, the stream's velocity dispersion (\S\ref{sec:gapFormation}), and realistic formation conditions \citep{overdensity1, overdensity2, bar1, reviewerBar} make this unlikely.

Analytic solutions have been found for five of the six profiles explored. Brief descriptions of each model are in Table \ref{table:models}. Further discussions and derivations of analytic velocity kicks can be found in Appendix \ref{sec:modelAppendix}. For the NFW case, our work is an extension and correction\footnote{In \cite{reviewerNFW}, there is a sign error in the logarithm term in the $a<1$ (or $s<r_\mathrm{p}$) case, which we have confirmed with those authors.} of \cite{reviewerNFW}.
\begin{table}
\begin{center}
\begin{tabular}{c c c c c c c}
    \hline
    Model  & Density Profile & 
    Mass-Size relation & \\
    \hline
    Point Model    &   $\rho(r) = M_\mathrm{tot} \delta^3(r)$ & N/A  \\ 
    Shell Model     &      $\rho(r) = \frac{M_\mathrm{tot} \delta^2(r- R_\mathrm{halo})}{4\pi R_\mathrm{halo}^2}$     & $R_\mathrm{halo}$ generated in simulation \\
    Constant model      &     $\rho(r) = \frac{3M_\mathrm{tot} \Theta(R_\mathrm{halo}-r)}{4\pi R_\mathrm{halo}^3}$ & $R_\mathrm{halo}$ generated in simulation \\
    Plummer Model       &     $\rho(r) = \frac{3M_\mathrm{tot} r_\mathrm{p}^2}{4\pi (r+r_\mathrm{p})^{5/2}}$ & $ r_\mathrm{p} = 1.62\,\mathrm{kpc} \left(\frac{M_\mathrm{tot}}{10^8\mathrm{M}_\odot}\right)^{1/2}$\\
    
    NFW Profile       &    $\rho(r) = \frac{\rho_0  \Theta(R_\mathrm{halo, NFW}-r)}{\frac{r}{R_s}(1 + \frac{r}{R_s})^2}$ & $R_\mathrm{halo, NFW}$ in eqn.~(\ref{eq:NFWMass})  \\
    
    Truncated NFW Profile       &    N/A & $R_\mathrm{halo}$ generated in simulation \\
    \hline
\end{tabular}
\caption{Density profile and mass-size relation for each model employed. $\Theta(r)$ is the Heaviside step function.  More extended discussions of each model can be found in Appendix \ref{sec:modelAppendix}.  $R_\mathrm{halo}$ and $\rho_0$ are directly simulated for each subhalo, as described in \S \ref{sec:galacticus}. For the NFW profile, $R_\mathrm{halo, NFW}$ is set so that the total mass of this subhalo is $M_\mathrm{tot}$.}
\label{table:models}
\end{center}
\end{table}

\subsubsection{Orbits}\label{sec:orbits}

With velocity kicks successfully calculated, we will next determine how they modify the orbits of stream particles. Naturally, the stream will follow the orbit equation:
\begin{equation}
    \frac{\mathrm{d}^2 u}{\mathrm{d}\theta^2} + u = -\frac{1}{L^2} \frac{\mathrm{d}\Phi(u)}{\mathrm{d}u} \label{eq:orbit},
\end{equation}
where $u \equiv \frac{1}{r}$, and $\Phi$ is the gravitational potential.  Due to the impulse approximation, interactions happen too quickly to be considered an additional force and rather provide a discontinuous boundary condition at $t=0$.  Equation~(\ref{eq:orbit}) has a fixed point for a circular orbit:
\begin{equation}
    u_\mathrm{0} + \frac{1}{L^2} \frac{\mathrm{d}\Phi(u_\mathrm{0})}{\mathrm{d} u}= 0, \label{eq:circular}
\end{equation}
where $u_\mathrm{0}$ is the inverse radius for a circular orbit. According to Bertrand's theorem \citep{bertrand1, bertrand2}, there will be a closed-form for the unperturbed orbit, but not for the modified orbit in a non-Keplerian potential. \footnote{Bertrand's theorem proves that Keplerian and harmonic orbits are the only power-law potentials with closed-form solutions. Therefore, the unperturbed orbit is simply solvable just because it is circular, not because the central potential has changed. Bertrand's theorem technically does not apply to realistic dark matter potentials such as flat rotation curves, however no known solution exists for these for other reasons.} However, since we assume $\Delta v_y \ll v_\mathrm{str}$, we can expand our perturbation to first order:
\begin{equation}
    \frac{\mathrm{d}^2 \Delta u}{\mathrm{d}\theta^2} + \left(u_0 + \Delta u\right) = -\frac{1}{L_0^2}\left( \frac{\mathrm{d}\Phi(u_0)}{\mathrm{d} u} + \frac{\mathrm{d}^2\Phi(u_0)}{\mathrm{d} u^2}\Delta u  -2 \frac{\mathrm{d}\Phi(u_0)}{\mathrm{d} u} \frac{\Delta L}{L_0}\right),\label{eq:circularFirstOrder}
\end{equation}
where we have Taylor expanded both $L$ and $\Phi$. Cancelling the $0^\mathrm{th}$ order solution using eqn.~(\ref{eq:circular}), and using $\frac{\Delta L}{L_0} = \frac{r' \Delta v_y} {r' v_\mathrm{str}} =\frac{\Delta v_y} { v_\mathrm{str}}$, we find
\begin{equation}
    \frac{\mathrm{d}^2 \Delta u}{\mathrm{d}\theta^2} + \left( 1 +\frac{1}{L_0^2}\frac{\mathrm{d}^2\Phi(u_0)}{\mathrm{d} u^2} \right)\Delta u =   -2 u_0 \frac{\Delta v_y} { v_\mathrm{str}}. \label{eq:orbitFinal}
\end{equation}
Note that $\Delta L$ cannot depend on $\Delta v_x$ due to geometry, nor on $\Delta v_z$ because these contributions vanish to first order. When this framework is carried beyond first order in future work, stream-plane deformations in the $z$-axis may become a valuable observable. Denoting
\begin{equation}
\gamma\equiv \sqrt{1 +\frac{1}{L_0^2}\frac{\mathrm{d}^2\Phi(u_0)}{\mathrm{d} u^2} },
\end{equation}
we find 
\begin{equation}
    \Delta u(\theta) = \frac{-2 u_0 \Delta v_y}{v_\mathrm{str} \gamma^2} \left(1 - \cos\gamma\theta\right) - 
\frac{u_0\Delta v_x}{ v_\mathrm{str} \gamma} \sin\gamma\theta.\label{eq:deltaU}
\end{equation}
Initial conditions are fixed by $\Delta u(\theta=0)=0$, and $\frac{\mathrm{d}\Delta u(\theta=0)}{\mathrm{d}\theta} = \frac{-\Delta v_x}{L_0}$.

The parameter $\gamma$ describes the epicyclic motion of perturbed orbits in the potential. This is crucial for gap formation (which is the result of perturbations to a stream orbit, but less critical for the evolution of the subhalo population. Therefore, in this work, we take $\gamma =\sqrt{2}$, as appropriate for flat rotation curves, even though this is inconsistent with the NFW potential that we use to describe the Milky Way halo and within which our subhalos orbit. While this results in some inconsistency, it provides a more realistic potential for gap formation calculations (a similar choice was made by \citealt{erkal3}). We estimate $\gamma^2 \approx 2.6$ at $r_0$ in the NFW profile of our Milky Way halos. As the angular evolution of stream particles is inversely proportional to $\gamma$ (as will be shown in eqn.~\ref{eq:theta}) this would lead to a 15\% shift in angular evolution (and gap) size compared to our assumption of $\gamma^2=2$. In future work, we plan to incorporate baryons into the evolution of the subhalo population. We will then compute $\gamma(r)$ directly from the full host halo potential, including contributions from the baryons of the Milky Way.

Since angular momentum is conserved after the perturbation, we rewrite $L = r^2 \dot{\theta}$ as 
\begin{equation}
    \dot{\theta} = \left(L_0 + \Delta L \right)\left(u_0 + \Delta u\right)^2, \label{eq:thetaDotInitial}
\end{equation}
or, to first order,
\begin{equation}
    \dot{\theta} \approx 
    L_0 u_0^2 + \Delta Lu_0^2 + 2 L_0 u_0 \Delta u.
    \label{eq:thetaDot}
\end{equation}
Defining
\begin{equation}
c_L(y) = \left|\frac{v_\mathrm{str}}{r_0} +\frac{\Delta v_y(y)}{\gamma^2r_0}
\left( \gamma^2 -4\right)\right| ,
\end{equation}
and
\begin{equation}
c_\mathrm{o}(y) = \frac{2\sqrt{\gamma^2\Delta v_x^2(y) + 4\Delta v_y^2(y)}}{\gamma^2r_0},
\end{equation}
and substituting in $\Delta u(\theta)$ we can simplify eqn.~(\ref{eq:thetaDot}): 
\begin{equation}
    \frac{d\theta}{dt} = c_\mathrm{L} +  c_\mathrm{o}\cos(\gamma\theta+\alpha),\label{eq:thetaDotFinal}
\end{equation}
where trigonometric identities have been used, and 
\begin{equation}
\alpha(y) = \left\{\begin{array}{ll} \arctan\left(\frac{\gamma X(y)}{2 Y(y)}\right) & \hbox{for } \mathbf{\Delta v} \neq 0,\\
0 & \hbox{for } \mathbf{\Delta v} = 0
\end{array}\right.
\end{equation}
represents a phase shift onto the perturbed orbits.

Eqn.~(\ref{eq:thetaDotFinal}) is equivalent to  Eqn.~(12) of \cite{erkal1} and follows the same derivation, but it is a more intuitive and useful form. In this new form,  Eqn.~(\ref{eq:thetaDotFinal}) is exactly integrable.  Before solving for $\theta(t)$, however, it is worth summarizing the steps thus far.  In the perturbed orbit, there is a shifted component of  $\dot{\theta}$, $c_\mathrm{L}(y)$ that varies with distance from the impact, but is constant over the gap's evolution.\footnote{Formally, $c_L(y)=\frac{v_\mathrm{str}}{r_0} +\frac{\Delta v_y(y)}{\gamma^2r_0}
\left( \gamma^2 -4\right)$. However, in the case where $c_L<0$, meaning $\Delta v_y \gtrsim v_\mathrm{str}$, the first order approximations of eqns.~(\ref{eq:circularFirstOrder};\ref{eq:thetaDot}) breaks down.  This only occurs in an incredibly small fraction of subhalos, so treating $c_L \rightarrow |c_L|$ avoids singularities without meaningfully affecting gap statistics.}  A second component, governed by $c_\mathrm{o}(y)$, encapsulates the oscillations as the gap expands. Both $c_\mathrm{L}(y)$ and $c_\mathrm{o}(y)$ are time independent, and only depend on a stream-point's position at $t=0$.  This form makes it clear that the linear orbital shift is governed purely by $\Delta v_y$, while the oscillations involve both components of the velocity kicks on equal footing.  The phase shift, $\alpha(y)$, reflects the fact that the ``circular orbit'' of a stream-point has been shifted by $\Delta L(y)$, and thus the stream is immediately out of equilibrium at $t=0$, wherever $\Delta v_y(y) \neq 0$. 

Integrating eqn.~(\ref{eq:thetaDotFinal}),\footnote{Without loss of generality, we use circular symmetry to fix the lower limit: $\theta(t=0)=0$.} 
\begin{equation}
\begin{split}
 \int_0^\theta  \frac{1}{c_\mathrm{L} +  c_\mathrm{o}\cos(\gamma\theta'+\alpha)} \mathrm{d}\theta' = \int_0^t \mathrm{d} t'\label{eq:integral}
\end{split}
\end{equation}
can be exactly solved as:\footnote{Here, \cite{erkal1} performed a Taylor expansion---consistent with the prior first-order approximations---to solve this integral. However, as this integral can be solved analytically, we choose to do so here and avoid introducing any further approximation.}
\begin{equation}
\begin{split}
\theta(y,t)= \mathcal{C}(y, t)-\frac{\alpha(y)}{\gamma}+\\
\frac{2}{\gamma}\left\{ \begin{array}{ll} 
\tan^{-1}\left[\frac{\sqrt{c_L^2 - c_\mathrm{o}^2}}{c_L - c_\mathrm{o} }\tan\left(\frac{\gamma\sqrt{ c_L^2 - c_\mathrm{o}^2} \times t}{2} + \delta\right)\right]  & \hbox{for } c_\mathrm{L} > c_\mathrm{o},\\
 \tan^{-1}\left(\frac{\gamma c_\mathrm{L}t}{2} + \delta \right)  & \hbox{for } c_L = c_\mathrm{o},\\
 \tan^{-1}\left[\frac{\sqrt{c_\mathrm{o}^2-c_L^2}}{c_\mathrm{o} - c_L}\tanh\left(\frac{\gamma\sqrt{c_\mathrm{o}^2-c_L^2} \times t}{2}+ \delta\right)\right]  & \hbox{for } c_\mathrm{L} < c_\mathrm{o},
\end{array}\right.\label{eq:theta}
\end{split}
\end{equation}
where 

\begin{equation}
\delta(y) = \left\{ \begin{array}{ll} \tan^{-1} \left[\sqrt{\frac{c_L - c_\mathrm{o}}{c_L + c_\mathrm{o}}} \tan\left(\frac{\alpha}{2}\right)\right] & \hbox{for } c_L> c_\mathrm{o},\\
0  & \hbox{for } c_L= c_\mathrm{o},\\
\tanh^{-1} \left[\sqrt{\frac{c_\mathrm{o}- c_L }{c_\mathrm{o}+ c_L}} \tan\left(\frac{\alpha}{2}\right)\right] & \hbox{for } c_L < c_\mathrm{o},\\\label{eq:delta}
\end{array}\right.
\end{equation}
arises from the lower limit of integration and   
\begin{equation}
    \mathcal{C}(y, t) = \frac{2\pi}{\gamma} \mathrm{Floor}\left(\frac{\frac{\gamma \sqrt{\lvert c_L^2(y) - c_\mathrm{o}^2(y)\rvert}\times t}{2} + \delta(y)}{\pi + \frac{1}{2}} \right)
\end{equation}
represents the linear component of gap evolution.\footnote{Note that $\mathrm{Floor}(x)\equiv \lfloor x\rfloor \Theta(x)+ \lceil x\rceil\Theta(-x)$ is a modified floor function that acts as a traditional floor function for positive arguments, and ceiling function for negative arguments. It is equivalent to type converting to integers in most coding languages.}  Because $\tan^{-1}\left(a \tan(b)\right): \mathbb{R}\rightarrow \left[-\pi/2, \pi/2/\right]$ only maps to the principle branch of $\tan b$,  $|\theta(t)|$ could never exceed $\frac{\pi}{\gamma}$ without $\mathcal{C}(y,t)$, and we could not represent the unperturbed stream.\footnote{While it takes some work to show, eqn.~(\ref{eq:theta}) does reduce to $\theta(t)= v_\mathrm{str}t/r_0$ in the absence of interactions.}  Like $\mathbf{\Delta v}_\mathrm{NFW}$ in eqn.~(\ref{eq:NFWKick}), the $c_L(y) > c_\mathrm{o}(y)$ and $c_L(y) < c_\mathrm{o}(y)$ cases of eqns.~(\ref{eq:theta}--\ref{eq:delta})  are related by analytic continuation. The $c_L(y)=c_\mathrm{o}(y)$ cases can be derived L'H\^opital's rule on either the first or third cases.

Note that eqn.~(\ref{eq:theta}) is only exact when $c_L(y) \ll c_\mathrm{o}(y)$---that is, when $\mathrm{max}(\Delta v) \ll v_\mathrm{str}$, meaning the second and third cases of $\theta(y,t)$ exceed the validity of this first order framework. However, these stream-destroying gaps are extremely rare, and the framework denoted here is still more exact than that of previous work. In addition to non-sinusoidal oscillations in eqn.~(\ref{eq:theta}), we find an average speed of $v_\mathrm{avg}(y) = r_0 \sqrt{c_L^2 - c_\mathrm{o}^2}$, rather than $v_\mathrm{avg}=c_L r_0$ as predicted in \cite{erkal3}.  This suggests that stream velocity is affected by $\Delta v_x(y)$ in an exciting new way. However, not all second-order effects have been included, so this potential dependence on $\Delta v_x(y)$ should be seen as preliminary.

\subsubsection{Gap Formation}\label{sec:gapFormation}
$\theta(y, t)$ can be thought of as a transfer function, moving stream material around the galactic centre. The secular component, $\frac{v_\mathrm{str}t}{r_0}$, merely corresponds to uniform circular motion. The perturbed component,
\begin{equation}
\Delta \theta(y,t) \equiv\theta(y,t) - \frac{v_\mathrm{str}t}{r_0},
\end{equation}
will transport stars at a non-uniform rate, leading to the gap development. 

Considering a one-dimensional stream with no variation in density perpendicular to the stream, the density is solely a function of the angular coordinate $\theta$.  Similar to \cite{erkal1}, we define an initial density $\rho_0(\theta_0) = \rho_0 =$ constant\footnote{This is not, in general, accurate due to effects including the epicyclic oscillations of stream stars and non-circular stream orbits, and future work intends to model the rich dynamics of stream overdensities \citep{overdensity1, overdensity2}.} (where $\theta_0$ is the unperturbed angular position), and a final density (after the gap has formed) of $\rho(\theta)$ in the perturbed coordinate $\theta$. The continuity equation (i.e., conservation of mass) then requires:
\begin{equation}
    \rho(\theta) \mathrm{d} \theta = \rho_0(\theta_0) \mathrm{d}\theta_0.
\end{equation}
Using the relation $\theta = \theta_0 + \Delta \theta$ we find
\begin{equation}
     \rho(\theta) \left( 1 + \frac{\mathrm{d}\Delta\theta}{\mathrm{d}\theta_0} \right) = \rho_0 .
\end{equation}
Then, since $\theta_0 = y/r_0$,
\begin{equation}
    \rho(\theta) \left( 1 + r_0 \frac{\mathrm{d}\Delta\theta}{\mathrm{d}y} \right) = \rho_0.
\end{equation}

We can define a normalized density as $\tilde{\rho}(y, t) \equiv \frac{\rho(y, t)}{\rho_0}$, which, noting that $\rho(y,t) \equiv \rho(\theta(y,t))$, from the above, can be seen to be 
\begin{equation}
    \tilde{\rho}(y, t) = \frac{1}{1 + r_0 \frac{\mathrm{d}\Delta\theta(y, t)}{dy}}. \label{eq:f}
\end{equation}
Equation~(\ref{eq:f}) is solved numerically for each model.

Assuming a cylindrical stream of cross-sectional area $A$, the projected surface density is $\Sigma(\theta) \sim \rho(\theta) A / 2 \sqrt{A/\pi}$, proportional to the density. Therefore, $\tilde{\Sigma}$ (the surface density relative to the original surface density) is simply equal to $\tilde{\rho}$ (assuming no changes in stream structure perpendicular to the stream).

Stars in the stream will have some random motion, characterized by a velocity dispersion, $v_\mathrm{d}$.  This dispersion arises both from any intrinsic dispersion in the progenitor system of the stream and from tidal heating \citep{Penarrubia2019}. These random motions will, over time, cause a gap to be repopulated by stars, reducing the width and depth of the gap. This effect was not considered by \cite{erkal3}, but here, we employ a simple model to account for this ``refilling'' effect\footnote{Of course, \cite{erkal3} \emph{did} include a cut on kick velocity, excluding interactions which produced a kick below 0.1~km/s. These small kicks would be the ones most likely to be ``filled in'' by the refilling effect we consider here. Nevertheless, this process is likely to be important even for larger kick velocities given the dispersion in a typical stellar stream.}. For simplicity, we adopt a constant value of $v_\mathrm{d} = 1.2$ km/s for the Pal-5 stream. In addition to the ``dispersionless'' motion predicted above, a star will also (on average) travel an angular distance of $v_\mathrm{d} t/r_0$ over the lifetime, $t$, of its gap.  We therefore apply a Gaussian smoothing on this angular scale to the density along the stream to model this refilling of the gap due to the velocity dispersion of the constituent stars. For computational efficiency, only gaps with $\tilde{\rho}<0.9$ are kept.

The size of a gap, $G$, can only be defined with respect to a cut-off depth, $\tilde{\rho}_\mathrm{c}$. This cut-off will depend on the current observational data available for a given stream. Therefore
\begin{equation}
    G(\tilde{\rho}_\mathrm{c}, t) = \frac{\left(\mathrm{max}\{y |\tilde{\rho}(y, t) =\tilde{\rho_\mathrm{c}}\} -\mathrm{min}\{y |\tilde{\rho}(y, t) =\tilde{\rho}_\mathrm{c}\}\right)}{r_0},
    \label{eq:gapSize}
\end{equation}
assuming $\rho_\mathrm{min}(t) \leq \rho_\mathrm{c}$.

To count the predicted number of gaps in a stream as a function of cut-off depth, one must conversely introduce a cut-off gap size. \footnote{There are, in principle, multitudes of infinitesimally small gaps, but these are of no observable consequence.} In this work, we use a cut-off size of $1^\circ$ in line with \cite{erkal3}. Therefore, 
\begin{equation}
    \mathrm{Count}(\tilde{\rho}_\mathrm{c}) \equiv \sum_{i \in \{\mathrm{Timesteps}\}} \sum_{j \in \{\mathrm{Gaps}\}_i} \Theta\left(G_{ij}(t_i) > 1^\circ\right) w_{ij},
    \label{eq:count}
\end{equation}
where $\Theta$ is a Heaviside step function, and $w_{i,j}$ is the length-adjusted sub-sampling weight of a given subhalo defined in eqn.~(\ref{eq:weights}). 

\section{Results}\label{sec:results}

We now present statistical results from applying this model to the Pal-5 stream.  Pal-5, one of the first streams discovered, is taken to be 3.4~Gyr old \citep{pal5age}, 9~kpc long at the present time \citep{pal5length}, and at an average distance of 13~kpc \citep{pal5r} from the Galactic centre. In \S\ref{sec:properties}, we compare the subhalo population to the previous analytical work \citep{erkal3}.  In \S\ref{sec:sizes}, we then examine the predicted distribution of gap sizes, and in \S\ref{sec:count}, we predict the number of gaps per stream. 

\subsection{Subhalo Population Properties}\label{sec:properties}

\begin{figure}
\includegraphics[width=85mm]{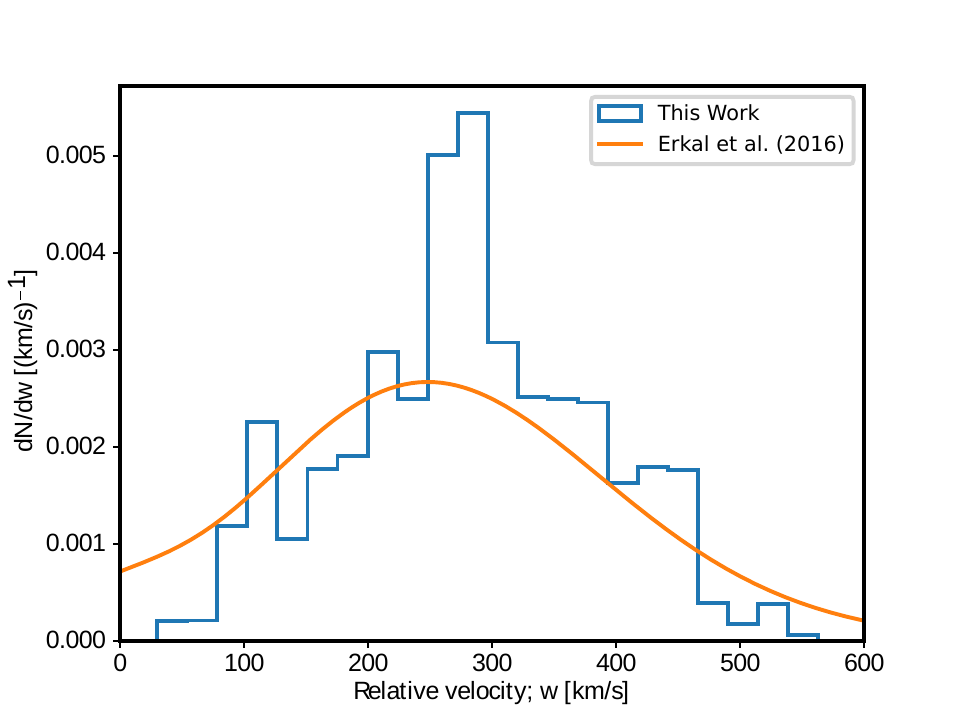}
\caption{The distribution of relative velocities between subhalos and the stream at redshift $z\approx0$. All infalling subhalos within 2.5~kpc of the stream were included. The orange line shows the distribution used by \protect\cite[][their equations 5--7]{erkal3}, which predict relative velocities of \emph{infalling} subhalos. The blue line shows results from this work, also for infalling subhalos, averaged over all realizations of the Milky Way halo.} 
\label{fig:w}
\end{figure}

\begin{figure}
\includegraphics[width=85mm]{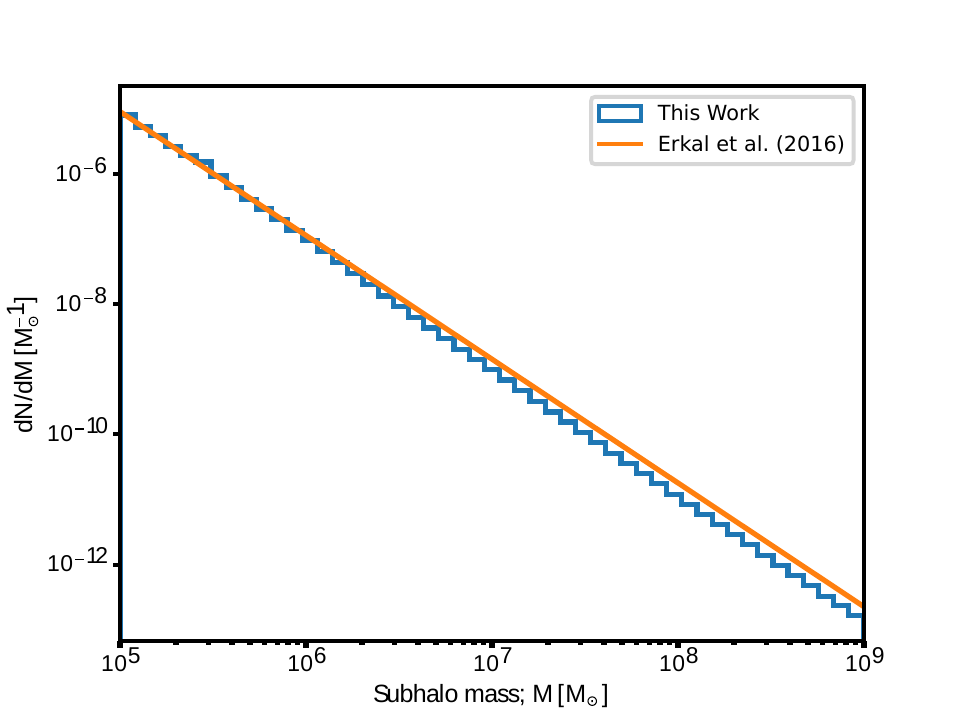}
\caption{The distribution of present-day subhalo bound masses (i.e. the subhalo mass function) at redshift $z=0$. All subhalos within the Milky Way halo are included. The orange line shows the distribution used by \protect\cite{erkal3}---which is a fit to the results of the Aquarius N-body simulation \citep{2008MNRAS.391.1685S}---while the blue line shows results from this work, averaged over all realizations of the Milky Way halo. We find a logarithmic slope of $-1.92$, very close to that found for the Aquarius simulations ($-1.9$; \citealt{2008MNRAS.391.1685S}).}
\label{fig:mass}
\end{figure}

\begin{figure}
\includegraphics[width=85mm]{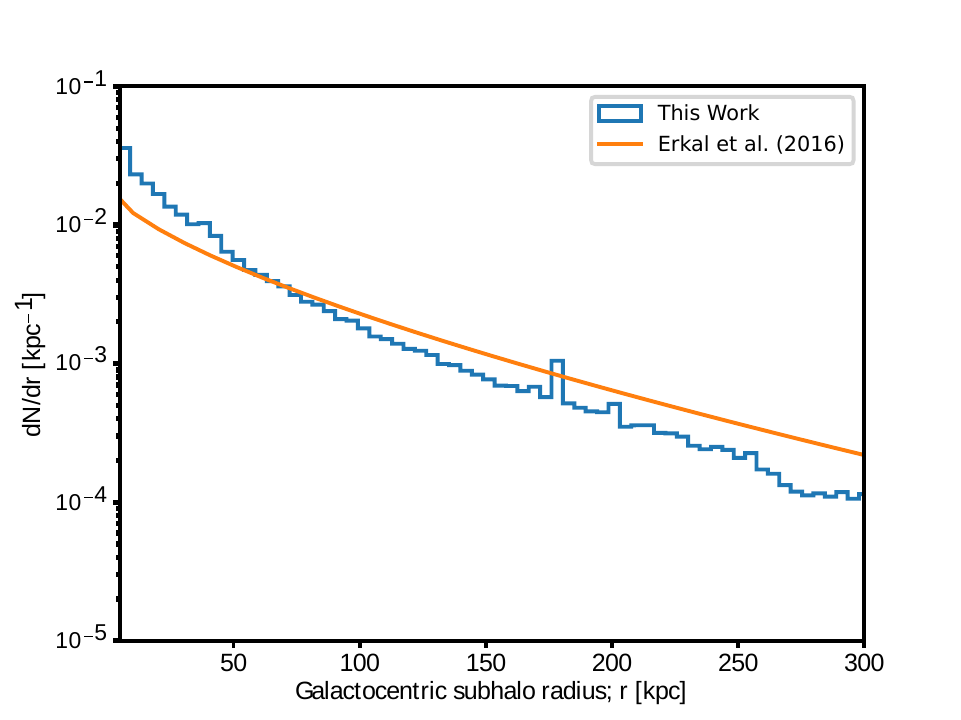}
\caption{The distribution of the galactocentric radii of subhalos at redshift $z=0$. All subhalos with bound masses $10^5 \mathrm{M}_\odot < M_\mathrm{b} < 10^9 \mathrm{M}_\odot$ were included. The orange line shows the distribution used by \protect\cite{erkal3}, while the blue line shows results from this work, averaged over all realizations of the Milky Way halo.}
\label{fig:r}
\end{figure}

Figures \ref{fig:w}--\ref{fig:r} show the position, mass, and velocity distributions of the subhalo population, comparing results from this work with those adopted by \cite{erkal3}. The comparison is made at $z=0$---in this work the subhalo population evolves over time, while in \cite{erkal3} these distributions did not evolve, but were fixed to measurements from N-body simulations at $z=0$.  Figure~\ref{fig:w} plots the distribution of relative velocities of the subhalos with respect to the stream, including only subhalos that are infalling and within 2.5~kpc of the stream (as was used by \cite{erkal3} in estimating this distribution). The selection of subhalos that are undergoing their closest approach with the stream results in a bias toward higher relative velocities. \cite{erkal3} account for this bias in constructing their distribution function. In this work, to match this bias, only subhalos that make their closest approach to the stream in the timestep being considered are counted. There is overall good agreement between the distribution predicted by {\sc Galacticus} and that adopted by \cite{erkal3}, although the distribution from {\sc Galacticus} is slightly less broad.

Next, Figure \ref{fig:mass} shows the distribution of subhalo masses, both from this work and \cite{erkal3}. All subhalos within the Milky Way halo are included (i.e., there is no selection on radial position or of infalling subhalos).  The results from this work agree very closely with those from \cite{erkal3}, as is to be expected as {\sc Galacticus} is known to accurately match subhalo mass functions from N-body simulations \citep{2020MNRAS.498.3902Y} to which \cite{erkal3} calibrated their model.

Finally, Figure \ref{fig:r} shows the radial distribution of subhalos with masses in the range $10^5 \mathrm{M}_\odot < M_\mathrm{b} < 10^9 \mathrm{M}_\odot$ (and without any selection of infalling subhalos). ``Spikes'' in the results from {\sc Galacticus} are due to random fluctuations enhanced by our sub-sampling approach. The overall shape predicted by {\sc Galacticus} is somewhat different from that adopted by \cite{erkal3}, with Galacticus predicting a subhalo population that is more centrally-concentrated than that used by \cite{erkal3}.

This difference in the shape of the radial distribution of subhalos as predicted by Galacticus and by N-body simulations has also been noted by \cite{2023ApJ...945..159N}, who compared Galacticus predictions matched to their ``\emph{Symphony}'' suite of zoom-in N-body simulations and showed that their Milky Way simulations predict 40--50\% fewer subhalos at small radii (i.e. the radii relevant for the Pal-5 orbit) than {\sc Galacticus}. {\sc Galacticus} was not calibrated to match the radial distribution of subhalos (instead being calibrated to match subhalo mass functions and the subhalo $V_\mathrm{max}$--$M$ relation; \citealt{2020MNRAS.498.3902Y})---further calibration in this statistic may improve the agreement with N-body simulations. However, as noted by \cite{2023ApJ...945..159N}, Galacticus' predictions agree with the \emph{Symphony} N-body results within the $2\sigma$ statistical uncertainties of those simulations. Furthermore, as also discussed by \cite{2023ApJ...945..159N}, subhalo radial distributions measured from N-body simulations may be affected by numerical issues, such as subhalo ``withering'' \citep{10.1093/mnras/stx2956} and artificial disruption \citep{10.1093/mnras/stab1215}. Improved subhalo finder algorithms \citep{2023arXiv230810926M}---which find  35--120\% more subhalos in the inner regions compared to the analysis of \cite{2023ApJ...945..159N}---and higher resolution simulations may help to resolve these issues in the near future, allowing for a more accurate assessment of the subhalo radial distribution.

\subsection{Gap Sizes}\label{sec:sizes}

\begin{figure}
\includegraphics[width=85mm]{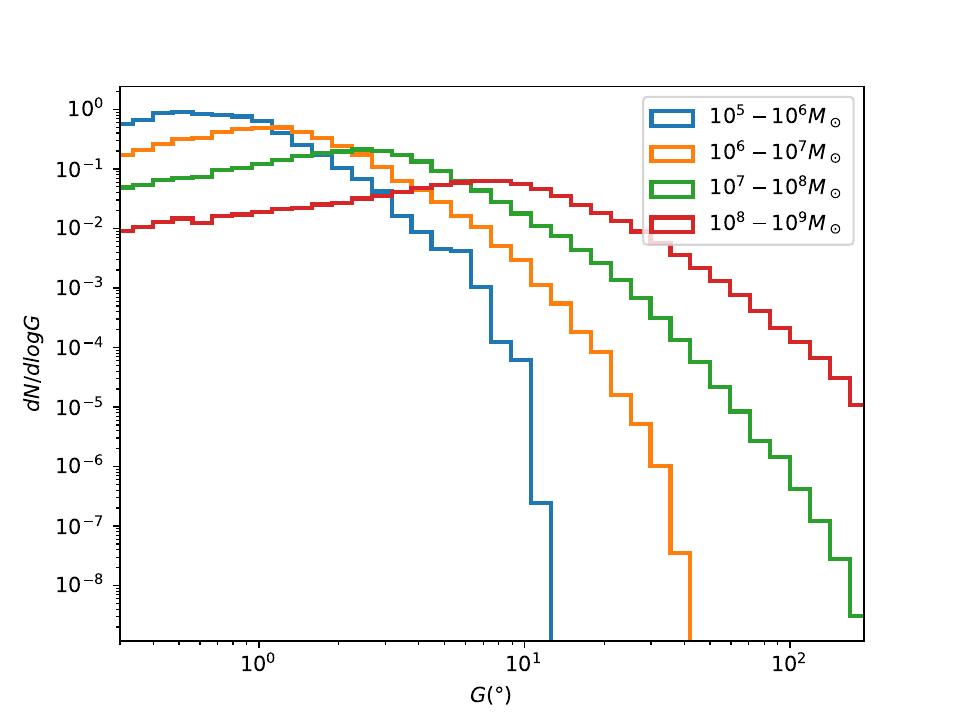}
\caption{The distribution of gap sizes, $G$, derived assuming tidally stripped NFW profiles, is shown for four intervals of subhalo mass from $10^5 \mathrm{M}_\odot$ to $10^9 \mathrm{M}_\odot$ as indicated in the figure. For this figure, gaps are defined using a threshold of $\tilde{\rho}_\mathrm{cut} = 0.5$.}
\label{fig:gapDecade}
\end{figure}

\begin{figure}
\includegraphics[width=85mm]{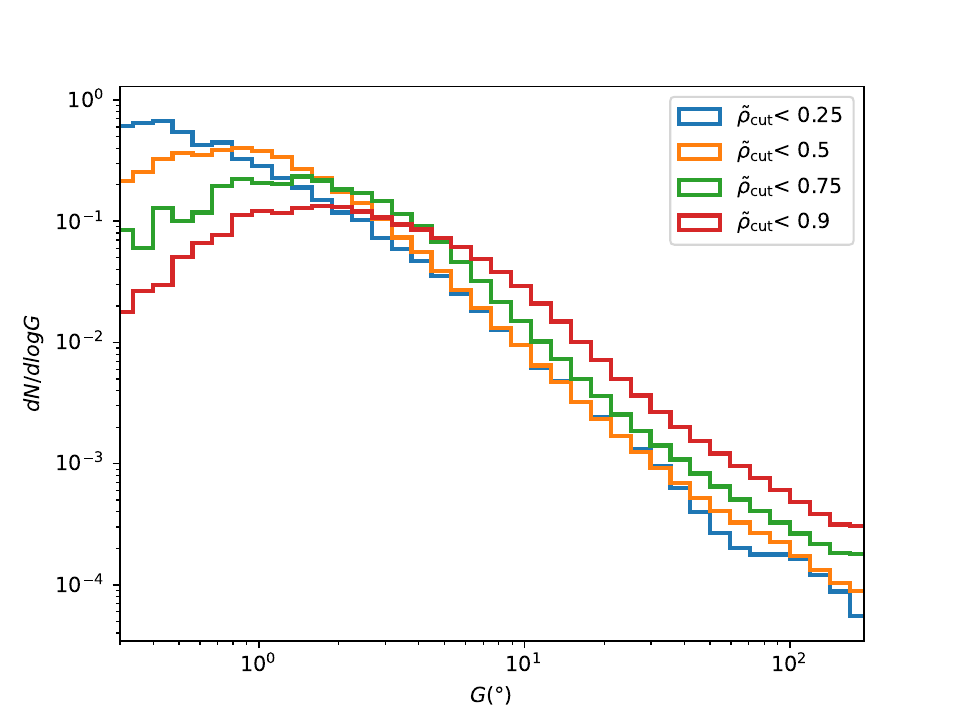}
\caption{The distribution of gap sizes, $G$, derived assuming tidally stripped NFW profiles, is shown for four different thresholds $\tilde{\rho}_\mathrm{cut}$ as indicated in the figure. Subhalos with masses in the range $10^5 \mathrm{M}_\odot < M_\mathrm{b} < 10^9 \mathrm{M}_\odot$ are included.}
\label{fig:gapMultipleCuts}
\end{figure}

\begin{figure*}
\begin{tabular}{cc}
\includegraphics[width=85mm]{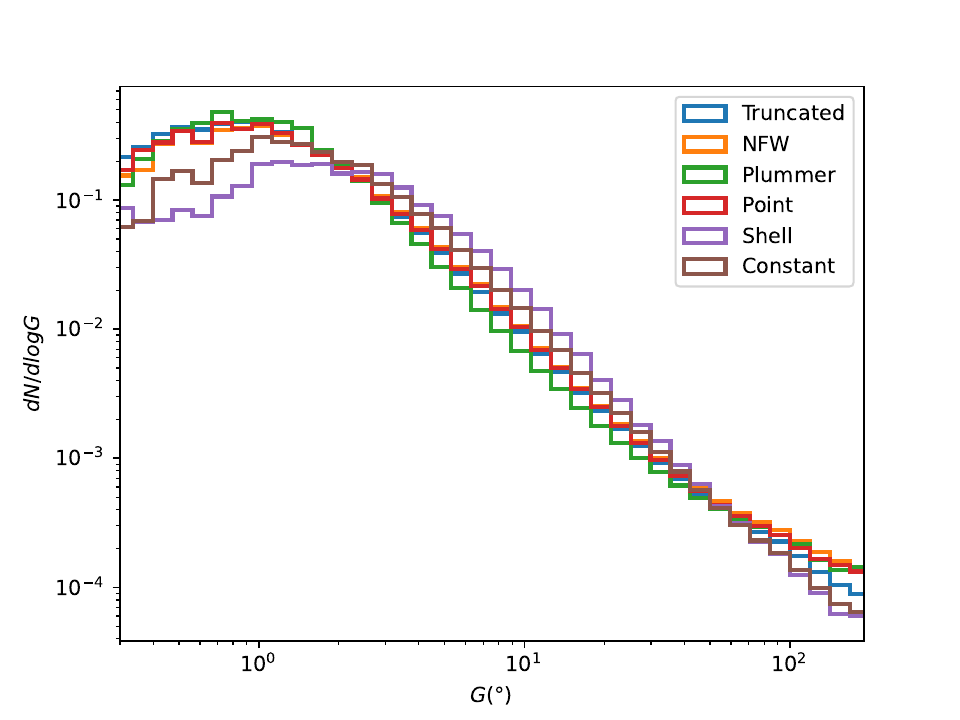} &
\includegraphics[width=85mm]{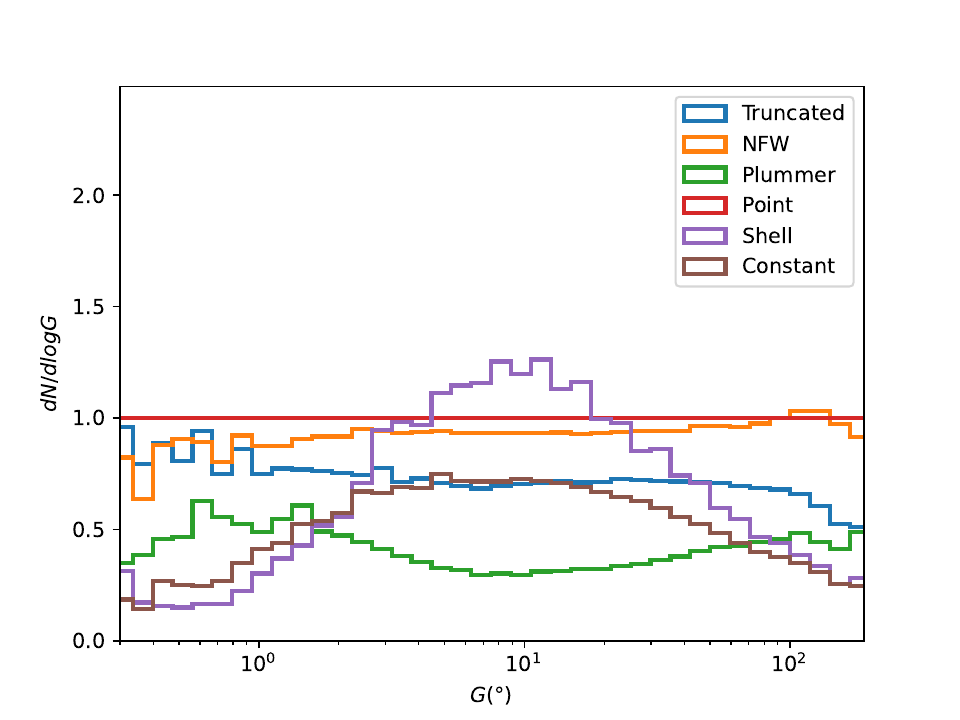}
\end{tabular}
\caption{The distribution of gap sizes, $G$, computed under different models for the dark matter subhalo density profile, as indicated in the figure.  Subhalos with masses in the range $10^5 \mathrm{M}_\odot < M_\mathrm{b} < 10^9 \mathrm{M}_\odot$ are included, and gaps are defined using a threshold of $\tilde{\rho}_\mathrm{cut} = 0.5$. The left panel shows the absolute gap size distribution, while the right panel shows results normalized to the number of gaps produced by a point model.}
\label{fig:gapAllModels}
\end{figure*}

Figures~\ref{fig:gapDecade}--\ref{fig:gapAllModels} present predicted distributions of gap sizes in the Pal-5 stream, generated using the model described in \S\ref{sec:methods}. Figure~\ref{fig:gapDecade} presents distributions split by subhalo mass, assuming tidally stripped NFW profiles, defined using a density threshold of $\tilde{\rho}_\mathrm{cut} = 0.5$. As subhalo masses increase, the distribution of gap sizes is shifted toward larger values, as expected (more massive subhalos cause stronger perturbations, leading to larger gaps).

Figure \ref{fig:gapMultipleCuts} presents similar distributions, but here, all subhalo masses are included in each curve. Each curve shows results for a different choice of density threshold, $\tilde{\rho}_\mathrm{cut}$, as indicated in the figure. For smaller values of $\tilde{\rho}_\mathrm{cut}$ (corresponding to the selection of deeper gaps), the distribution of gap sizes is shifted toward smaller values.

Finally, Figure \ref{fig:gapAllModels} presents normalized gap size distributions for different choices of dark matter subhalo density profile in the left panel, and the number of gaps relative to a point-mass perturber in the right panel. In all cases, a density threshold of $\tilde{\rho}_\mathrm{cut} = 0.5$ is used, and contributions from subhalos over the entire mass range ($10^5 \mathrm{M}_\odot < M_\mathrm{b} < 10^9 \mathrm{M}_\odot$) are included. Looking at the left panel, despite relatively distributions between profiles ($<1\%$ of subhalo-stream interactions have impact parameters small enough to probe the inner structure of the subhalo), there is a clear excess of gaps in the 3-$10^\circ$ size interval when using an NFW or tidally stripped NFW profile relative to when a Plummer profile is used.  While it is only a factor of  $\sim 2$ difference, this excess is significant when using streams as a sensitive dark matter probe.  In at the right panel, we see relative agreement between the Point, and NFW models. The Truncated model agreed reasonably well with these two for small gaps ($< 1^\circ$), but increasingly predicts fewer gaps as size increases. The Shell model, surprisingly, predicts the most $\sim10^\circ$ gaps, outperforming even the most-dense point mass. The constant density model mirrors the shape of the Shell model, yet predicts fewer gaps at all sizes above $2^\circ$.  Also unexpectedly, the Plummer model exhibits an inverse trend compared to the other non-cuspy perturbers, with the fewest gaps around $10^\circ$, and peaks at $0.4^\circ$ and $100^\circ$. It appears uniquely bad at simulating realistic dark matter.

\subsection{Number of Predicted Gaps}\label{sec:count}

\begin{figure}
\includegraphics[width=85mm]{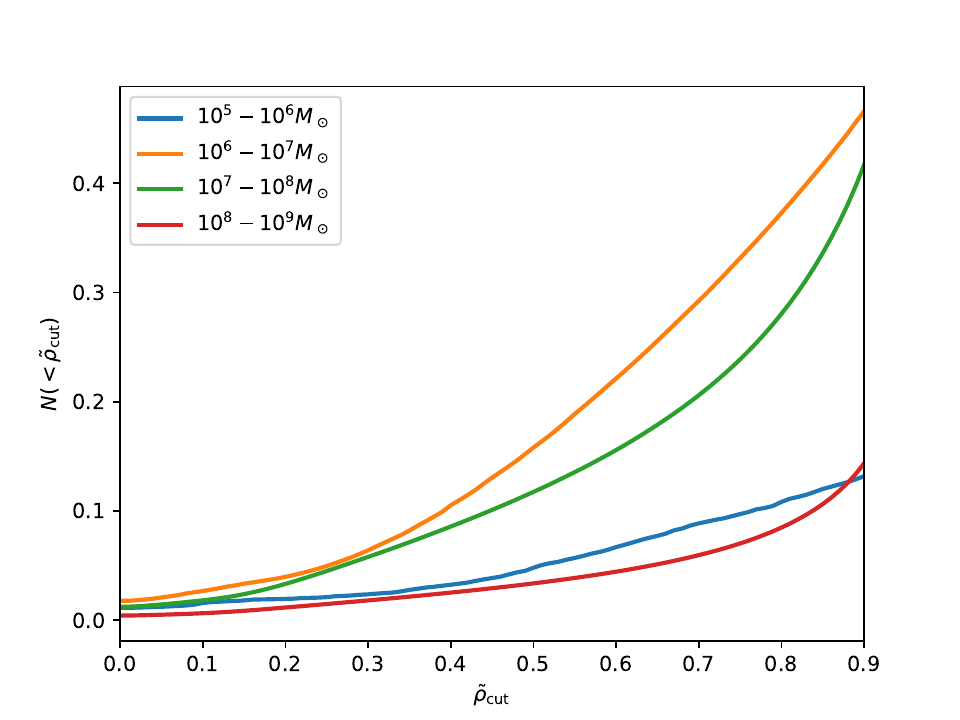}
\caption{The mean number of gaps of size greater than $1^\circ$ per Pal-5 stream as a function of the threshold, $\tilde{\rho}_\mathrm{cut}$, originating from encounters with subhalos in different mass intervals (as indicated in the figure).}
\label{fig:countDecade}
\end{figure}

\begin{figure}
\includegraphics[width=85mm]{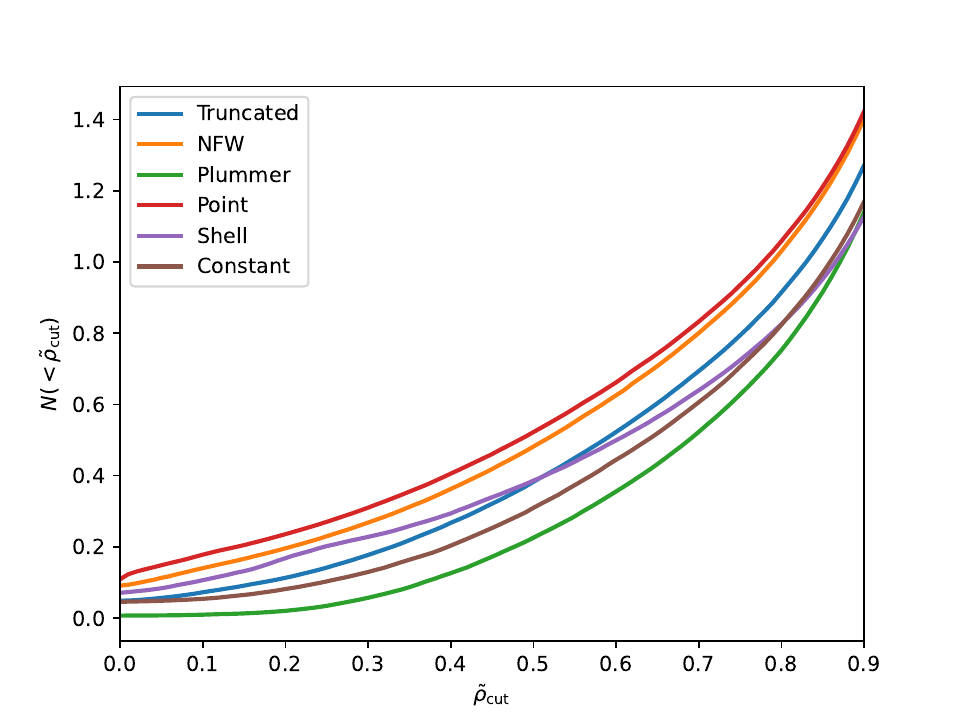}
\caption{The mean number of gaps of size greater than $1^\circ$ per Pal-5 stream as a function of the threshold, $\tilde{\rho}_\mathrm{cut}$, is shown for six different models of the subhalo density profile. Subhalos across the entire mass range considered ($10^5 \mathrm{M}_\odot < M_\mathrm{b} < 10^9 \mathrm{M}_\odot$) are included.}
\label{fig:countAllModels}
\end{figure}

\begin{table}
\begin{center}
\begin{tabular}{c c c c c c c}
    \hline
    $\tilde{\rho}_\mathrm{cut}$  & 0.10  & 0.20  & 0.30  & 0.50  & 0.75  & 0.90  \\
    \hline
    Truncated         & 0.072 & 0.113 & 0.177 & 0.382 & 0.794 & 1.271 \\ 
    NFW               & 0.140 & 0.196 & 0.268 & 0.481 & 0.904 & 1.402 \\
    Plummer           & 0.010 & 0.020 & 0.057 & 0.225 & 0.627 & 1.139 \\
    Point             & 0.178 & 0.236 & 0.310 & 0.522 & 0.935 & 1.422  \\
    Shell             & 0.106 & 0.168 & 0.228 & 0.386 & 0.724 & 1.127  \\
    Constant              & 0.054 & 0.081 & 0.129 & 0.309 & 0.705 & 1.169  \\
    \cite{erkal3} & ---   & ---   & ---   & 0.300 & 0.700 & ---   \\
    \hline
\end{tabular}
\caption{The mean number of gaps of size greater than $1^\circ$ per Pal-5 stream is shown for different thresholds, $\tilde{\rho}_\mathrm{cut}$ (columns), and for different choices for the subhalo density profile (rows), and including contributions from subhalos across the entire mass range of $10^5 \mathrm{M}_\odot < M_\mathrm{b} < 10^9 \mathrm{M}_\odot$. In the final row, we show results quoted by \protect\cite{erkal3} where available.}
\label{table:count}
\end{center}
\end{table}

\begin{figure}
\includegraphics[width=85mm]{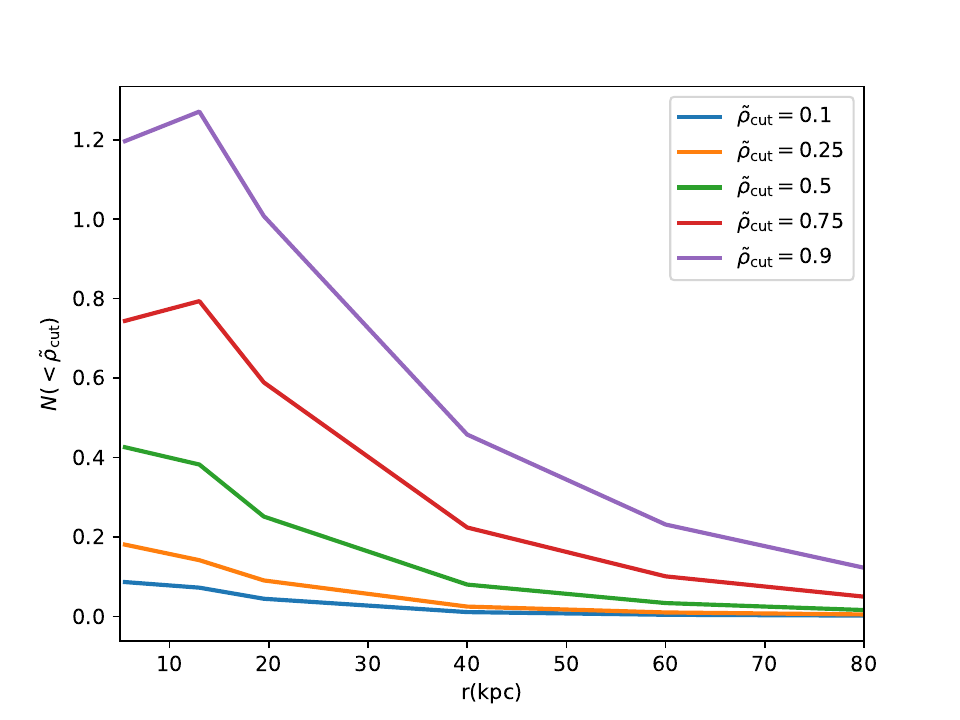}
\caption{The mean number of gaps of size greater than $1^\circ$ per Pal-5 stream as a function of stream radius is shown for a variety of the $\tilde{\rho}_\mathrm{cut}$ threshold values.  Subhalos across the entire mass range considered ($10^5 \mathrm{M}_\odot < M_\mathrm{b} < 10^9 \mathrm{M}_\odot$) are included, and $\gamma = \sqrt{2}$, $v_\mathrm{stream}=220$ kpc are used at all radii for consistency.}
\label{fig:countRadii}
\end{figure}

Figures~\ref{fig:countDecade}--\ref{fig:countAllModels} show the predicted mean number of gaps of size greater than $1^\circ$ per Pal-5 stream as a function of gap depth threshold, $\tilde{\rho}_\mathrm{cut}$.  As previously stated, the subhalo population is not constant in time (as in \citealt{erkal3}), but evolves alongside the host halo.  Also previously mentioned, the number of gaps in Pal-5 depends on the minimum depth observable by current photometric and spectroscopic data.  When shallower gaps are included, the total number of gaps naturally increases. We choose $\tilde{\rho}_\mathrm{cut}=0.5$ as a representative value, but results can be computed for any choice of threshold. In both figures, we assume tidally stripped NFW density profiles for subhalos.

In Figure~\ref{fig:countDecade}, the mean number of gaps is split into contributions from four intervals of subhalo mass, as indicated in the figure. The number of gaps contributed per decade of subhalo mass peaks in the $10^6 \mathrm{M}_\odot$ to $10^7 \mathrm{M}_\odot$ interval. For higher masses, the number of gaps contributed is reduced simply because these higher mass subhalos are far less abundant, while for lower masses, the number of gaps is reduced because, while there are many such subhalos, interactions with sufficiently small impact parameter to generate wide and deep gaps become exceedingly rare.

Figure~\ref{fig:countAllModels} also shows the predicted mean number of gaps $> 1^\circ$ per Pal-5 stream as a function of gap depth threshold, $\tilde{\rho}_\mathrm{cut}$, but now including contributions from subhalos of all masses. Individual lines correspond to different choices for the subhalo density profile, as indicated in the figure. The cuspy NFW and tidally stripped NFW profiles produce more gaps (of given depth and size) than the cored Plummer profile due to their more concentrated mass distributions. The point mass naturally produces the most gaps, as it is the most concentrated mass distribution, and the constant density model produces fewer gaps than the point and NFW-like profiles. It is counterintuitive that the constant density subhalo produces more gaps than the Plummer model. This is not fully understood but suggests that gaps may be very sensitive to halo size. Most counterintuitively, the shell profile produces more gaps than every profile except a point mass and an NFW profile. The interactions are strictly smaller for the shell than for any other profile; however, gap formation may follow a more complex relationship. This is discussed further in \S\ref{sec:discussion}.  The tidal truncation and heating of the NFW profiles causes a modest, $\sim 8$\% reduction in the total number of gaps produced.

These same results are tabulated in Table~\ref{table:count} for a few values of $\tilde{\rho}_\mathrm{cut}$. We also quote the results given by \cite{erkal3} where available. These should be compared with our results for a Plummer profile---there is close agreement between our results and those of \cite{erkal3} in these cases. The differences (which are at the level of 20\%) arise from the differences in subhalo population statistics (see \S\ref{sec:properties}), the evolution of the subhalo population over time (included in our model, but not in that of \citealt{erkal3}) and, to a much lesser degree, our use of numerical solutions to eqn.~(\ref{eq:thetaDotInitial}) instead of a series expansion solution.

Figure \ref{fig:countRadii} shows how gap counts vary as a function of the stream's galactocentric radius.  Stream length is left fixed at all radii, along with and $\gamma = \sqrt{2}$, and $v_\mathrm{str}=220$ for consistency.  At all depths, gaps are more common in the Milky Way's inner region but can still be found in streams on larger orbits.  This difference is the most noticeable for the shallowest gaps, and nearly disappears at $\tilde{\rho}_c = 0.1$, however the cause is poorly understood.  To our knowledge, this is the first time gap predictions have explicitly been modeled as a function of distance. Further investigation into the differing changes between shallow and deep gaps would be of value.

\section{Discussion}\label{sec:discussion}

The data presented in \S\ref{sec:results} represent exciting new results.  Even for the Plummer model, Table \ref{table:count} shows a  $\sim10-30\%$ difference in gaps (of $1^\circ$ or larger) between this work and \cite{erkal3}.  The updated radial distribution of subhalos shown in Figure \ref{fig:r} predicts more subhalos near Pal-5, yet we find fewer gaps in this work. Therefore, this difference demonstrates the need for a semi-analytic approach in realistic predictions.  Comparing our Plummer and Truncated profiles in Table~\ref{table:count} and Figure~\ref{fig:countAllModels}, the Plummer model gives $40\%$  fewer gaps at  $\tilde{\rho}_c=0.5$, and $7$ times fewer gaps at $\tilde{\rho}_c=0.1$.  Therefore, the proper mass profile is vital for realistic gap modelling. 

Although some authors have moved to a power spectrum approach \citep{streamsPowerSpectrum, gaps10}, if the underlying mass profile and subhalo population are inaccurate, these systematic errors will still be present.  While \cite{Adams2024} uses a realistic subhalo population, there may still be significant biases due to the Plummer profile employed. Moreover, this work only uses 400 host halos,\footnote{There are an additional 185 host halos in their ``high mass'' dataset.  However, this population has a different host halo mass and cannot be combined statistically.} while we find that $1,600,000$ host halos are required to achieve a $1\%$ statistical accuracy.  It is possible that aspects of their methodology lead to greater accuracy-per-realization, particularly by not employing subsampling; however, in this work, uncertainties are found to be Poissonian. So, even assuming that a lack of subsampling makes their realizations 10 $\times$ more efficient, a sample size of 400 host halos would have uncertainties roughly of $\sqrt{1.6\times 10^6/400 /10}\times 1\% \sim 20\%$.

While not all mass profiles employed are plausible models of dark matter, they are all deeply informative. For instance, it is unclear why the Plummer model produces so few gaps (Figure \ref{fig:countAllModels}) and why they are all so small (Figure \ref{fig:gapAllModels}). Given that the derivatives of velocity kicks generate gaps,  perhaps the smoothing at small radii makes the Plummer model uniquely bad for simulating gap formation. Some candidates, such as Self-Interacting Dark Matter (SIDM), have cored subhalo profiles; will they produce similarly few deep gaps? The constant density profile is effectively one big core, and yet it still produces $\sim5$ times more deep gaps than the Plummer model. All data in Table \ref{table:count} uses the same subhalo population, and outside of the subhalo radius, all models are equivalent to a point mass. Therefore, it is both significant and poorly understood why two non-cuspy profiles produce such different results. Without further testing, it is impossible to determine which model more accurately approximates cored SIDM subhalos. 

Another mystery is the shell profile. This model strictly has the smallest velocity kicks of all bounded profiles, a fact that we have verified directly. However, in Figure~\ref{fig:countAllModels}, it produces more shallow gaps than the Plummer and constant density models and produces more deep gaps than all profiles besides Point and NFW. Looking at Figure \ref{fig:gapMultipleCuts}, we see that gap depth and width are inversely related. Therefore, one possible explanation is that the shell profile creates expectationally many narrow gaps, just large enough to pass our $>1^\circ$ cutoff. If this were true, it would suggest that gap size is not a uniquely valuable metric; perhaps the slope of a gap and the total displaced material can reveal more information. 

A second possibility is related to the presence of the derivative in eqn.~(\ref{eq:f}).  Gaps are implicitly dependent on $\frac{d\mathbf{\Delta v}}{d\theta}$, rather than $\mathbf{\Delta v}$. Therefore, the discontinuity in $M(r)_\mathrm{shell}$ at the halo radius may translate to a spike in stream density. Looking directly at the derivative of eqn.~(\ref{eq:velocitySpikes}), we see a discontinuity at the subhalo radius:

\begin{equation}
\begin{split}
    \frac{d\Delta v_\mathrm{i, shell}(\theta)}{d\theta}= -\frac{2\mathrm{G}M_\mathrm{tot}r_0y\wperp^2X_i}{w \rmin^2}\times \\
    \begin{cases}
    \left(\frac{-1}{\rmin^2} + \frac{\sqrt{R_\mathrm{halo}^2 -  \rmin^2}}{\rmin^2R_\mathrm{halo}} - \frac{1}{R_\mathrm{halo}\sqrt{R_\mathrm{halo}^2 - \rmin^2}}\right) & \hbox{for } \rmin(\theta) < R_\mathrm{halo},\\
    \left(\frac{-1}{\rmin^2} \right) & \hbox{for } \rmin(\theta) \geq R_\mathrm{halo}.\\\label{eq:velocitySpikes}
    \end{cases}
\end{split}
\end{equation}

However, all bounded mass distributions will have this same singular term, so this behaviour should not be unique to the Shell profile. This result is hard to reconcile with intuitions of smoothly varying integrals, but if this second possibility is correct, it further suggests non-trivial ``fingerprints'' due to subhalo shape. 

Before this work, it seemed unlikely that subhalos lighter than $10^5 M_\odot$ could contribute significantly to gap formation. In Figures 15--16 of \cite{erkal3}, the $10^5 M_\odot-10^6 M_\odot$ mass decade consistently made the fewest gaps. However, Figure \ref{fig:countDecade} shows that with a cuspy mass profile, $10^5-10^6 M_\odot$ subhalos outperform $10^8-10^9 M_\odot$ subhalos at all gap depths, and rival $10^7-10^8 M_\odot$ subhalos for small $\tilde{\rho}_c$.  Further work is required to place a lower bound on gap-producing subhalos, but it is very possible that perturbers  $\lesssim 10^4M\odot$ can be detected.  This would be an unprecedentedly deep test of deviations from the CDM subhalo mass function.

Looking at Figure \ref{fig:countDecade}, it is surprising that the $10^5$--$10^6 M_\odot$ decade creates deep gaps. Naively, one would expect the numerous, light subhalos to create only shallow gaps, but a non-trivial fraction may have slow relative velocities or impact parameters.  Subhalos with $w=0$ have a galactocentric velocity of $220$ km/s, and it can be seen in Figure~\ref{fig:w} that slow subhalos are not infrequent.  Furthermore, if $\wperp= \frac{w_\perp}{w}$ is also small, these gaps will have close interactions over a wide region of the stream. Figure \ref{fig:gapDecade} shows that the $10^5-10^6 M_\odot$  decade has the smallest gaps, so high-quality observational data such as is expected from Via will be crucial in their detection.  Mass profiles only create a small variation in gap size (Figure \ref{fig:gapAllModels}).

Despite being similar in their inner region, the Truncated profile predicts $35\%$ fewer gaps than the NFW profile at $\tilde{\rho}_c = 0.2$, and $50\%$ fewer gaps at $\tilde{\rho}_c = 0.1$. Deep gaps are thought to be caused by deep impacts, but perhaps other forces are at play. The NFW velocity kicks found in \S\ref{sec:NFW} are a better model than the Plummer profile, but they should not be considered a realistic approximation of CDM subhalos. 

Figure \ref{fig:countRadii} presents another new result. While there are more gaps predicted in the inner region of the galaxy, streams outside of the visible galaxy will experience fewer gaps from baryonic effects and the LMC (Large Magellanic Cloud). While few streams in the outer galactic region have been observed, this is expected to change through upcoming surveys. 

Statistically, the results presented in this work are \emph{precise} to approximately $1.0\%$ given the number of realizations (and rotations) of subhalo populations used. This has been verified by repeating calculations using different random seeds at various sample sizes.  However, the \emph{accuracy} of the model, which systematic biases may limit, is much more difficult to assess. Plausibly, the most significant systematic bias may arise from the lack of inclusion of baryons in our model.\footnote{Note that we fix the parameter $\gamma$ in eqn.~(\protect\ref{eq:deltaU}) to $\gamma^2 =2$ appropriate to the flat rotation curve of the actual Milky Way (i.e., including baryons).} In particular, the contribution of the Milky Way galaxy to the gravitational potential and tidal field could have significant consequences for the population of subhalos in the inner regions of the halo \citep{2017MNRAS.471.1709G,2017MNRAS.469.1997D,10.1093/mnras/stx2238,Nadler_2018,Nadler_2021}. Beyond this effect, other systematic biases likely remain in our model due to its assumption of spherical symmetry in the subhalo population, the use of a circular orbit for the stream, and the fact that we average over all possible formation histories of a Milky Way-mass halo, rather than only over those consistent with constraints on the formation history of the actual Milky Way (e.g., the presence of the Large Magellanic Cloud is uncommon in Milky Way-mass halos, but is known to affect the dynamics of stellar streams; \citealt{Shipp_2021}).

One remaining chief question is, 'Which gaps are observable?' This, of course, is stream-specific and will change through the upcoming spectroscopic surveys of the next decade. Moreover, real streams do not have a continuum of infinitesimal member particles that can be arbitrarily rearranged. Several studies have attempted to address this question for the case of Pal-5 \citep{gapSize1,gapSize2,gapSize3}, but none give estimates of detectability as a function of gap depth and size. \cite{2025arXiv250207781L} quantify the minimum detectable subhalo mass for a variety of streams, finding that, for the case of Pal-5, gaps are sensitive to subhalo down to $10^7\mathrm{M}_\odot$ using a combination of GAIA and DESI data. A crucial missing ingredient required to bring models of the type presented here into contact with experimental data is detailed studies that characterize the detectability of gaps as a function of their size and depth. With such studies in hand, detailed forecasts could be made for the ability of stellar streams to discriminate between different dark matter models. Excluding the Plummer profile, Table \ref{table:count} indicates a $19-150\%$ spread in the number of $>1^\circ$ gaps between profiles (varying with depth). Therefore, it is highly likely that realistic mass profiles can be discriminated using stellar streams.

\section{Conclusion}\label{sec:conclusions}

Stellar streams are a powerful probe of dark matter, but due to the low number of gaps in individual streams, their true value lies in the statistical signal averaged over the full ensemble of streams.  Analytical models are not accurate enough, and it is unlikely that N-body simulations can be run at a sufficient scale. Therefore, semi-analytic modeling may be the only path towards large-scale tests of streams, perhaps the only viable hope of truly solving dark matter (see arguments in \S{\ref{sec:introduction}}). At present, predictions  $\sim 250$ known streams could be run in roughly $400,000$ CPU hours, which is moderately feasible. For analytic velocity kicks, this would take $25,000$ CPU hours, $\sim 3000$ times faster than the procedure of \cite{erkal3}.  Future improvements may further speed up non-analytic computation times by 10- to 100-fold, providing viability in a future epoch with thousands of streams.

While the approach described in \S\ref{sec:methods} has been indirectly validated to simulations through \cite{erkal1}, numerous improvements must still be made. Baryons must be modelled in subhalo evolution, and streams must be given realistic orbits, =non-uniform progenitors \citep{reviewerProgenitor}, and epicyclic overdensities must be included \citep{overdensity1, overdensity2}. However, none of these jeopardize the core drivers of simulation speed, described in Appendices \ref{sec:impactAppendix}-\ref{sec:hypergeometricAppendix}.  Baryonic perturbers such as Giant Molecular Clouds \citep{gmc}, the galactic bar \citep{bar1, reviewerBar}, the Large Magellanic Cloud  (LMC; \cite{Shipp_2021, reviewerLMC1, reviewerLMC2}), and spiral arms must be included \citep{reviewerArms}, but eqn.~(\ref{eq:integralFinal}) generalizes to all impulsive perturbers. Large perturbers, such as the LMC and the bar, affect streams over long periods of time, and their interactions cannot be treated as instantaneous.  These perturbers will likely need to be integrated directly, as opposed to fast, light perturbers that are largely impulsive.  Fortunately, there are only small numbers of these larger perturbers, meaning that integrating their effects directly is unlikely to add a huge computational burden.

While making a semi-analytical framework realistic will likely require a substantial effort, there is a secondary benefit, namely the ability of these approaches to be used in forward modeling pipelines.  The search for, say, the subhalo that caused the spur in GD-1 can tell us a great deal about a single perturber.  However, unless this one observation is enough to constrain dark matter uniquely, we must combine multiple streams of evidence.  Constructing likelihood functions in these cases can be impractical. Forward modeling opens the possibility of likelihood-free statistical approaches such as approximate Bayesian computing \citep{2020MNRAS.491.6077G, reviewerABC} and simulation based inference \citep{reviewerSBI}, as has been used in other areas of dark matter inference.

While this paper does not attempt to provide a detailed framework for such forward modeling, we feel this is one of the important challenges facing stream research and that improving our semi-analytic models can lead to key advances and insights in this area. For instance, the Shell profile (\S\ref{sec:shell}) indicates that gap width and depth are insufficient to fully characterize the strength of interaction  (see discussion in \S\ref{sec:discussion}). Perhaps the slope of a gap, $\mathrm{d}\tilde{\rho}(\theta)/\mathrm{d}\theta$, or the total ``volume'' of the underdense region,
\begin{equation}
    \int_{\tilde{\rho}^{-1}_\mathrm{left}(1)}^{\tilde{\rho}^{-1}_\mathrm{right}(1)} \tilde{\rho}(y')\mathrm{d}y',
    \end{equation}
are better observables to measure. Why do the pure-NFW profile and the tidally heated version (\S\ref{sec:NFW}; Table \ref{table:count}) have differing numbers of deep gaps, despite only differing in their outer regions? Given the non-analyticity of the stripped profile, one can only directly investigate this directly by modeling gaps in different scenarios and speculating on the results. However, this cause may be related to the surprising differences between gaps from Shell, Plummer, and constant density models, which can be directly compared (\S \ref{sec:plummer}-\ref{sec:constant}).  Simplified mass profiles are not adequate for realistic predictions, but can provide valuable insights. This is not to say that we should abandon particle spray models. Instead, these two approaches can work in tandem: closed-form results motivate simulations, and those simulations refine the approximations we use.

Eqn.~(\ref{eq:integralFinal}) demonstrates that velocity kicks are separable into geometric and profile-specific components. While this may not generalize to non-circular perturbers on generic orbits, the need for fast modelling leads us to understand in what regimes impulsive, first-order perturbations remain valid approximations. Given the sensitivity to subhalo profile (\S\ref{sec:discussion}), gaps may be affected by halo shape as well. Most calculations have assumed spherical halos, yet this is not generically true \citep{randomWalk}, and self-interacting subhalos will have different shapes due to isotropization. While one can simulate flybys of varying triaxiality for a variety of scenarios and tabulate these results, a more fruitful approach may be to expand subhalo mass distributions in multipoles. The dominant spherical mode will still obey eqn.~(\ref{eq:integralFinal}), and the effects of each higher moment on gaps could be assessed. Extending the may reveal candidate-specific modes of oscillation. 

The ultimate goal of this line of research is a perturber-dependent ``footprint'' on streams.  If gaps generated by a $10^5 M_\odot$ globular clusters \citep{globularGaps} can be separated from those of a $10^7 M_\odot$ NFW-like subhalo, most or all baryonic gap sources could be ignored. However, such an idea is likely optimistic, and should not be expected to be viable at this point \citep{reviewerFolding}.  Infall effects will certainly produce degeneracies and obfuscate any gravitational ``fossil record'' embedded in the stream. However, Table \ref{table:count} shows pervasive model dependence even after Gaussian smoothing. Therefore, there is every reason to believe that these various predictions can be unified into a more coherent framework.

\emph{If} such a framework can be found, and \emph{if} dark matter can be constrained to a specific point in parameter space, the nature of dark matter will still not be a solved problem. Even if dark matter is proved to be self-interacting with a specific cross-section $\sigma(v)$-- this will not uniquely identify a Lagrangian. However, even in the worst case, where deviations from CDM cannot be found at any astrophysical scale, this will still represent a major step forward. The vast majority of current dark matter candidates deviate from CDM on scales more massive than $10^5 M_\odot$ \citep{annika2}.  Eliminating these allows the direct-detection community to concentrate on more fruitful ideas, perhaps leading to a breakthrough.  However, before any of this can happen, gap predictions must be compared between CDM and other candidates.  The differences in predicted gaps represent the underlying power of this test. If this difference is too small to be observed in $\gtrsim 10,000$ streams, then it is unlikely that gaps can constrain dark matter. This will be the focus of the next paper in this series.

\section*{Acknowledgments}

We thank Ana Bonaca, Denis Erkal, and the entire Streams 2023 conference for inspiration, references, and clarifying discussions. We also thank Ethan Nadler and Sachi Weerasooriya for helpful comments that improved the clarity of this 
work.

\section*{Data Availability}

The data underlying this article will be shared on reasonable request to the corresponding author.


\bibliographystyle{mnras}
\bibliography{sample}
\appendix

\section{Symmetries and Computation of impact parameter}\label{sec:impactAppendix}
When calculating the impact parameter, $b$, it is exceedingly computationally intensive to directly minimize eqn.~(\ref{eq:distanceFinal}) for each subhalo. However, it is possible to pre-screen $\sim 98\%$ of the subhalos through various symmetries and physical bounds.  These filters, taking effectively no compute time themselves, naturally speed up this procedure by roughly a factor of 10. While this is not the computational bottleneck for a single subhalo, given the number of subhalos considered at this stage, the total computing time is significant.   For the remaining subhalos, the impact parameter can be found via polynomial roots, which is significantly faster and more precise than a global minimization procedure. By substituting equation 6 into equation A1

The first two filters are immediately applied to the output of {\sc Galacticus}, reducing the total memory requirements of the simulation. As discussed in \S\ref{sec:methods}, we keep only subhalos with 
\begin{equation}0 \leq t_\mathrm{min} \leq \frac{T_\mathrm{stream}}{N_t},\label{eq:streamImpact}\end{equation}
where $N_t$ is the number of timesteps in this simulation, and $t_\mathrm{min}$ is defined relative to the start of this timestep. While $t_\mathrm{min}$ depends on the precise location of the impact, subhalos that do not approach \emph{any} point along the stream in this timestep can be removed. Specifically, by substituting in eqn.~(\ref{eq:impactTime}), one finds that subhalos failing to satisfy
\begin{equation}
      -\frac{v T_\mathrm{stream}}{N_t} - r_\mathrm{stream} \leq \mathbf{r}' \cdot \hat{\mathbf{v}} \leq r_\mathrm{stream}. \label{eq:firstFilter}
\end{equation}
are removed. These algebraic inequalities take a negligible amount of computation time.

A second filter eliminates all subhalos that create insufficiently deep gaps. Again, we cannot calculate the specific gap depth without knowing the precise impact parameter and location. However, consistent with \cite{erkal3},  we find that all significant gaps have $\mathrm{halo}(|\Delta v_y|) \geq 0.1$~km/s.  Therefore, if $\mathrm{max}(|\Delta v_y|) < 0.1$~km/s \emph{everywhere} along the stream, then this subhalo can be discarded. We treat all mass profiles as point masses for this filter, as this provides simple results that can only overestimate interaction strength and, therefore, provide conservative bounds. Following \S\ref{sec:gaps}, removes all subhalos with 
\begin{equation}
    \frac{2\mathrm{G}M}{(v_\mathrm{sub}- v_\mathrm{stream})(r_\mathrm{sub, min} - r_\mathrm{stream})} < 0.1 \mathrm{km/s}, \label{eq:streamKick}
\end{equation}
where $(v_\mathrm{sub}, v_\mathrm{stream})$ are the respective galactocentric velocities of the subhalos, and  $(r_\mathrm{sub, min}, r_\mathrm{str})$ are respectively the (minimum) galactocentric position of the subhalo and the stream's circular radius.\footnote{Note that in this approximation, any subhalo with pericenter smaller than  $ r_\mathrm{stream}$ automatically passes, as there will be some point where $r_\mathrm{sub}(t) = r_\mathrm{str}$. This artifact disappears  when the stream is fixed to a specific orbital plane in \S\ref{sec:preparations}.}  Unlike \cite{erkal3} and \cite{Adams2024}, we do not place an upper limit on $b$, and thus this cleaning is crucial in removing unwanted subhalos.

Once the orbital plane of the stream has been fixed, we can repeat these constraints in a stronger fashion. $v_z$ is fixed, so it is natural to work in cylindrical coordinates. For the impact time, only subhalos with
\begin{equation}
      -\frac{v T_\mathrm{stream}}{N_t} - r_\mathrm{stream}\hat{v}_\rho \leq \mathbf{r} \cdot \hat{\mathbf{v}} \leq r_\mathrm{stream}\hat{v}_\rho,\label{eq:impactMiddle}
\end{equation}
are kept, where $\hat{v}_\rho \equiv \frac{\sqrt{v_x^2 + v_y^2}}{v}$. Similarly, all subhalos with 
\begin{equation}
    \frac{2\mathrm{G} M}{\sqrt{(v_\rho - v_\mathrm{stream})^2 + v_z^2} \sqrt{\left(r_\rho - r_\mathrm{stream}\right)^2 + r_z^2}} < 0.1 \mathrm{km/s}. \label{eq:kickMiddle}
\end{equation}
cannot create a sufficiently deep gap and are removed. $r_\rho$ and $r_z$ are respectively the planar and $z$ components of the $\mathbf{r}_\mathrm{pericenter}$.  These filters remove roughly two-thirds of all subhalos.

Finally, one can easily check whether there are \emph{any} local minima in the allowable timestep. However, to do so, we will first have to map our questions to a more usable form. Local minima and maxima of eqn~(\ref{eq:distanceFinal}) are found at 

\begin{equation}
   \frac{1}{2} \frac{d (d^2_\mathrm{min}(\phi))}{d\phi} =  
   -\mathbf{ \dot{r'}}(\phi)\cdot\left(\mathbf{r} +\mathbf{\Delta r}(\phi) \cdot \hat{\mathbf{v}}\right) =0, \label{eq:minAngle}
\end{equation}
which assumes $\mathbf{r}'(\phi)\cdot \dot{\mathbf{r}}'(\phi) =0$ for a circular stream orbit.

As discussed above, once we fix the stream to the $x$--$y$ plane, we can parameterize it as 
\begin{equation}
\mathbf{r}'(\phi) = r' \left(\cos\phi, \sin\phi, 0\right).\label{eq:rPrime}
\end{equation}
Plugging in, we find
\begin{equation}
    \left(r_x -\hat{v}_x \mathbf{r}\cdot\hat{\mathbf{v}} \right)\sin\phi -
    r_y\cos\phi +
    r'\hat{v}_x^2\cos\phi\sin\phi =0, \label{eq:minimizeWithAngles}
\end{equation}
or 
\begin{equation}
    c_x\sin\phi - c_y\cos\phi + d\cos\phi\sin\phi =0, \label{eq:minimizeWithCoefficients}
\end{equation}
with constants
\begin{equation}
c_x = r_x -\hat{v}_x \mathbf{r}\cdot\hat{\mathbf{v}}, \ \ \   
c_y = r_y, \ \ \ 
d = r'\hat{v}_x^2.
\end{equation}
This can be further simplified using Weierstrass substitution identities, 
\begin{equation}
    \sin\phi =\frac{2v}{1+v^2}, \ \ \ \cos\phi  =\frac{1-v^2}{1+v^2}, 
\end{equation}
leading to 
\begin{equation}
   f(v) = v^4 + b_3v^3 + b_1v - 1 = 0, \label{eq:minimizeWithPolynomials}
\end{equation}
with 
\begin{equation}
b_3 = \frac{2(c_x -d)}{c_y},\ \ \ b_1 = \frac{2(c_x + d)}{c_y}.
\end{equation}

While quartic equations can be directly solved, this will not generalize to non-circular orbits. In preparation for future work, we have adopted a different approach, which starts with an additional set of filters.  Using Sturm Chains \citep{sturm}, we can remove any subhalos without a local minimum in the region
\begin{equation}
    R = \left[\tan\left(\frac{\phi_{\mathrm{min}}}{2}\right), \tan\left(\frac{\phi_{\mathrm{max}}}{2}\right)\right] \cup  \left[\tan\left(\pi - \frac{\phi_{\mathrm{max}}}{2}\right), \tan\left(\pi - \frac{\phi_{\mathrm{min}}}{2}\right)\right],
\end{equation} 
with 
\begin{equation}
    \phi_{\mathrm{min}} = \arccos\left(\frac{\mathbf{r}\cdot \hat{\mathbf{v}}+ T v}{r' v_\rho}\right), \phi_{\mathrm{max}} = \arccos\left(\frac{\mathbf{r}\cdot \hat{\mathbf{v}}}{r' v_\rho}\right).
\end{equation} 
Again, the second region exists because $\arccos(v)$ only spans $[0, \pi)$, whereas the stream is defined between $[0, 2\pi)$ in this model. To find all turning points in the region $R$, the Sturm polynomials are 
\begin{equation}
\begin{split}
 P_0(v) = f(v) = v^4 + b_3v^3 + b_1v - 1, \\ 
 P_1(v) = f'(v) = 4v^3 + 3b_3v^2 + b_1, \\ 
 P_2(v) = -\mathrm{Rem}(P_0, P_1) = \frac{3b_3}{16}v^2 - \frac{3 b_1}{4} v + \left( \frac{b_1 b_3}{16} + 1\right), \\
 P_3(v) = 
-\left(\frac{64b_1^2}{b_3^4} - \frac{64}{3b_3^2} + \frac{32 b_1}{3b_3}\right) v + \left(\frac{256b_1^2}{3b_3^4} + \frac{16b_1^2}{3b_3^3} + \frac{16}{b_3}   \right),\\
P_4(v) = -\frac{b_3^4 \left(256 + 27 b_1^4 + 192 b_1 b_3 - 6 b_1^2 b_3^2 + 4 b_1^3 b_3^3 + 27 b_3^4\right)}{64 \left((-2 b_3^2 + b_1 (6 b_1 + b_3^3))^2\right)}.
\end{split}
\end{equation}

Subhalos are removed if the number of sign variations is constant across both regions. This guarantees that a turning point exists in the valid region, but it does not guarantee that this is a local minimum. While we cannot assess whether a turning point is a minima without knowing its precise location, we can determine whether there is any part of $R$ with $f'(v) <0$. Using a similar Sturm procedure, the polynomials for $f'(v)$ are  

\begin{equation}
\begin{split}
\tilde{P}_0 = f'(v) = 4v^3 + 3 b_3 v^2 + b_1,\\
\tilde{P}_1 = f''(v) = 12 v^2 + 6 b_3 v,\\
\tilde{P}_2 = - \mathrm{Rem}(\tilde{P}_0, \tilde{P}_1) = \frac{b_3^2}{2}v -b_1,\\
\tilde{P}_3 = \frac{-12 b_1}{b_3}\left(\frac{4 b_1}{b_3^3} + 1\right).
\end{split}
\end{equation}
Subhalos are kept only if the number of sign variations changes across either region or $f'(v)<0$ at the endpoint of either region. 

To summarize, we have filtered out $\sim 98\%$ of subhalos before calculating impact parameters. Now, the roots of eqn.~(\ref{eq:minimizeWithPolynomials}) are calculated numerically, which is insignificant to the total computation time. Because $f(v)$ is quartic, there will be either 0, 2, or 4 real roots. However, after noting that  $f(0)=-1$ and  $f(\pm\infty)= \infty$, we realize there are always at least 2 roots, reflecting the fact that all flybys will have a point of closest approach. These roots are converted back to angles with $\phi = \arctan{2v} $. These angles are plugged back into eqn.~(\ref{eq:distanceFinal}), and we define $b= \mathrm{min}(d_\mathrm{min}(\{\phi_\mathrm{turning}\}))$. Unlike global minimum finders, these solutions are exact, which is crucial for both keeping the proper subhalos and for correctly modeling the deepest, most informative impacts. The impact time is computed from eqn.~(\ref{eq:impactTime}), and impacts outside of this timestep are discarded. Relative velocities are then computed as in \S\ref{sec:preparations}, and we are ready to simulate gap formation.

\section{Derivation of New Impact Parameter Equation}\label{sec:velocityKickAppendix}
As discussed in \S\ref{sec:kicks}, eqns.~(\ref{eq:integralFinal}--\ref{eq:componentsFinal}) represent a powerful new formula for velocity kicks. This appendix derives these results, starting from eqn.~(\ref{eq:kickStartingPoint}). 

First, one can separate the two terms of this integral as
\begin{equation}
\begin{split}
\Delta v_i= c_1\int_{-\infty}^{\infty} \frac{-\mathrm{G} M(r)}{\left((y + w_\parallel t)^2 + w_\perp ^2 t^2 +b^2\right)^{3/2}}  \mathrm{d}t +\\
c_2\int_{-\infty}^{\infty} \frac{-\mathrm{G} M(r)t}{\left((y + w_\parallel t)^2 + w_\perp ^2 t^2 +b^2\right)^{3/2} }  \mathrm{d}t. \label{eq:kickTime}
\end{split}\
\end{equation} 
These terms are not manifestly even or odd in time because of a cross term in the denominator.  However, the cross term $2 y w_\parallel t$ vanishes at $y=0$, leaving an even $c_1$ term and an odd $c_2$ term. This simplification reflects our choices of $t$ and $y$, both of which are defined relative to the time and location of the impact. However, gravitational interactions are fundamentally local, and so a coordinate system unique to each $y$ point is needed. The simplest approach is to integrate over distance rather than time. Manifestly, 
\begin{equation}
r(t) = \sqrt{(y + w_\parallel t)^2 + w_\perp ^2 t^2 +b^2}.
\end{equation}
Inverting, one finds
\begin{equation}
\begin{split}
t_\pm(r) = t_\mathrm{off} \pm \frac{\sqrt{r^2 -\rmin^2}}{w},\label{eq:t}
\end{split}
\end{equation}
with
\begin{equation}
    \toff(y) \equiv \frac{-y\tilde{w}_\|}{w},\ \  \rmin(y) \equiv \sqrt{b^2 + y^2 \wperp^2},\ \  \tilde{w}_{\parallel, \perp} \equiv \frac{w_{\parallel, \perp}}{w}.
\end{equation}

Note that $t(r)$ is not single-valued. There is a negative branch while the subhalo is infalling and a positive branch after the scattering event. This is reflected in the Jacobian
\begin{equation}
\mathrm{d}t = \frac{\pm r}{w \sqrt{r^2 - \rmin^2}} \mathrm{d}r,  \label{eq:dt}  
\end{equation}
which is formally infinite at $\rmin$, the point of closest approach for a given point $y$. This singularity is a coordinate artifact that merely reflects that $r'(t_\mathrm{off}) =0$. The integral itself is well-defined. Looking specifically at the $c_1$ term, we  decompose the integral into infalling and outfalling parts:
\begin{equation}
\begin{split}
I_1 = c_1\left(\int_{-\infty}^{\tmin} \frac{-\mathrm{G} M(r)}{ r^3(t)}  \mathrm{d}t + 
\int_{\tmin}^{\infty} \frac{-\mathrm{G} M(r)}{ r^3(t)}  \mathrm{d}t\right)\\
=c_1\left(\int_{\infty}^{\rmin} \frac{\mathrm{G} M(r)}{w r^2 \sqrt{r^2 - \rmin^2}}  \mathrm{d}r + 
\int_{\rmin}^{\infty} \frac{-\mathrm{G} M(r)}{ wr^2 \sqrt{r^2 - \rmin^2}}  \mathrm{d}r\right).
\end{split}
\end{equation} 
Simplifying, one finds
\begin{equation}
    I_1= c_1\int_{\rmin}^{\infty} \frac{-2\mathrm{G} M(r)}{wr^2 \sqrt{r^2 - \rmin^2}}  \mathrm{d}r \label{eq:firstTerm}.
\end{equation}
Similarly, the $c_2$ term can be rewritten as 
\begin{equation}
\begin{split}
I_2 = c_2\left(\int_{-\infty}^{\tmin} \frac{-\mathrm{G} M(r)t}{ r^3(t)}  \mathrm{d}t + 
\int_{\tmin}^{\infty} \frac{-\mathrm{G} M(r)t}{ r^3(t)}  \mathrm{d}t\right)\\
=c_2\int_{\infty}^{\rmin} \frac{\mathrm{G} M(r)}{w r^2 \sqrt{r^2 - \rmin^2}}\left(t_\mathrm{off} - \frac{\sqrt{r^2 -\rmin^2}}{w}\right)  \mathrm{d}r + \\
c_2\int_{\rmin}^{\infty} \frac{-\mathrm{G} M(r)}{ wr^2 \sqrt{r^2 - \rmin^2}}\left(t_\mathrm{off} + \frac{\sqrt{r^2 -\rmin^2}}{w}\right)  \mathrm{d}r.
\end{split}
\end{equation} 
The more second term in each integral cancels, leaving 
\begin{equation}
    I_2= c_2\int_{\rmin}^{\infty} \frac{-2t_\mathrm{off}\mathrm{G} M(r)}{wr^2 \sqrt{r^2 - \rmin^2}}  \mathrm{d}r \label{eq:secondTerm}.
\end{equation}

Therefore, there was always a hidden time symmetry in the velocity kicks that has been made manifest.  Each component of the velocity kick is of the form,
\begin{equation}
    \Delta v_i =\left(c_1 + c_2 \toff\right)\int_{\rmin}^{\infty} \frac{-2\mathrm{G} M(r)}{w r^2 \sqrt{r^2 - \rmin^2}}  \mathrm{d}r.\label{eq:velocityKickFinalAppendix}
\end{equation}

Converting to a dimensionless distance variable $\tilde{r} \equiv r/\rmin$, and calling this prefactor
\begin{equation}
   X_i(y) \equiv c_1 + c_2 \toff(y),
\end{equation}
where explicitly
\begin{eqnarray}
X(y) &\equiv& b_x + w_x\toff(y) \nonumber \\
Y(y) &\equiv& = y \frac{w^2_\perp}{w^2} = y \wperp^2 \nonumber \\
Z(y) &\equiv& b_z + w_z\toff(y), \label{eq:componentsFinalAppendix}
\end{eqnarray}
we can simplify eqn.~(\ref{eq:velocityKickFinalAppendix}) further: 
\begin{equation}
    \Delta v_i(y) = \frac{-2\mathrm{G}X_i(y)}{w \rmin^2}\left(\int_{1}^{\infty} \frac{M(\tilde{r})}{ \tilde{r}^2 \sqrt{\tilde{r}^2- 1}}  \mathrm{d}\tilde{r} \right).
\end{equation}
And, since any bounded, isotropic mass distribution is indistinguishable from a point mass outside of its cutoff radius, we define $\tilde{R}_\mathrm{max} = {\mathrm{min}\left(R_\mathrm{sub}/\rmin,1\right)}$. Then,
\begin{equation}
    \Delta v_i(y) = \frac{-2\mathrm{G}X_i(y)}{w \rmin^2}\left(\int_{1}^{\tilde{R}_\mathrm{max}} \frac{M(\tilde{r})}{ \tilde{r}^2 \sqrt{\tilde{r}^2- 1}}  \mathrm{d}\tilde{r} +\int_{\tilde{R}_\mathrm{max}}^\infty  \frac{\tilde{M}_\mathrm{tot}}{ \tilde{r}^2 \sqrt{\tilde{r}^2- 1}}  \mathrm{d}\tilde{r}  \right).
\end{equation}

That is, the first term integrates the unique inner region of each mass profile,\footnote{Assuming the impact is deep enough for this term to contribute.} and the second term uses a point mass  at all larger distances. The point mass can be directly integrated. Writing $\tilde{M}(\tilde{r}) \equiv \frac{M(\tilde{r})}{M_\mathrm{tot}}$, we are left with  
\begin{equation}
    \Delta v_i = \frac{-2\mathrm{G}M_\mathrm{tot}X_i(y)}{w \rmin^2}\left( 1 - \frac{\sqrt{\tilde{R}_\mathrm{max}^2- 1}}{\tilde{R}_\mathrm{max}} + 
    \int_{1}^{\tilde{R}_\mathrm{max}} \frac{\tilde{M}(\tilde{r})}{ \tilde{r}^2 \sqrt{\tilde{r}^2- 1}}  \mathrm{d}\tilde{r} \right). \label{eq:integralFinalAppendix}
\end{equation}

This form of the velocity kicks exploits hidden underlying symmetries to reveal essential physics. Because only one integral needs to be performed per stream-point, and because the remaining integral is manifestly symmetric under $y\rightarrow -y$,  eqn.~(\ref{eq:integralFinalAppendix}) is $\sim 6$ times faster to numerically integrate than eqn.~(\ref{eq:kickStartingPoint}). Further speedups can be found in Appendix \ref{sec:hypergeometricAppendix}.

\section{Solutions for Integrable velocity kicks} \label{sec:modelAppendix}
When evaluating eqn.~(\ref{eq:integralFinalAppendix}), six different mass profiles, $M(\tilde{r})$ are used. In five of these cases, the velocity kicks can be evaluated analytically. The NFW profile is one such case; to the authors' knowledge, this analytic solution for the resulting velocity kicks has not been previously found.  While no analytic solution exists for the truncated NFW profile, additional details about that model can be found in Appendix \ref{sec:hypergeometricAppendix}.

\subsection{Point Mass}
The simplest possible model of a subhalo is a point mass.  While this overpredicts gap formation,\footnote{See \S\ref{sec:results}.} it was historically the first mass profile explored in \cite{yoon} and \cite{carlberg3}, and it is the best model for building simple intuition. There is no internal structure, and thus, 
\begin{equation}
    \Delta v_\mathrm{i, point}(y) \equiv - \frac{2\mathrm{G}M_\mathrm{tot}X_i}{w \rmin^2}.\label{eq:point}
\end{equation}

\subsection{Plummer Sphere}\label{sec:plummer}
The Plummer sphere is a useful toy model that has been utilized in numerous papers, including \cite{erkal1}, \cite{erkal2}, \cite{erkal3}, and \cite{Adams2024}. Its main advantage is the ease of analytical solutions\footnote{As discussed, the NFW profile is also integrable. However, this has not been shown in previous work.}. However, the cored inner region leads to an under-prediction of deep gaps. 

The profile is defined by the potential
\begin{equation}
\Phi_\mathrm{Plummer}(r) = - \frac{\mathrm{G} M_\mathrm{tot}}{\sqrt{r^2 + r_\mathrm{p}^2}},  \label{eq:Plummer}
\end{equation}
where $M_\mathrm{tot}$ is the total bound mass of the object, and the Plummer radius, $r_\mathrm{p}$, acts as a smoothing parameter.  At small distances ($r\lesssim r_\mathrm{p}$), $r_\mathrm{p}$ avoids numerical singularities, while at large distances ($r\gg r_\mathrm{p}$), this potential approaches the form of a point mass. During a flyby, the stream interacts only with $M_\mathrm{enc}(r)$, which has the form 
\begin{equation}
M_\mathrm{enc}(r) = M_\mathrm{tot} \frac{r^3}{\left(r^2 + r_\mathrm{p}^2\right)^{3/2}}.
\end{equation}

For comparison with \cite{erkal3} we adopt their choice of
\begin{equation}
r_\mathrm{p} = 1.62\,\mathrm{kpc} \left(\frac{M_\mathrm{tot}}{10^8\mathrm{M}_\odot}\right)^{1/2}
\end{equation}
when using Plummer profiles.  Because the Plummer model is not formally bounded, it does not follow eqn.~(\ref{eq:integralFinal}). Instead, one finds the simpler result
\begin{equation}
    \Delta v_\mathrm{i, Plummer}(y) = -\frac{2\mathrm{G}M_\mathrm{tot}X_i}{w\left(\rmin^2 + r_\mathrm{p}^2\right)} \label{eq:IPlummer},
\end{equation}
which reduces to a point mass as $r_\mathrm{p}\rightarrow 0$.  Although our nomenclature differs slightly, our results are equivalent to those of \cite[][and subsequent work]{erkal1}.

\subsection{Shell Model}\label{sec:shell}
Next, a shell model concentrates all mass at a radius of $R_\mathrm{halo}$: 
\begin{equation}
    M(r) = \begin{cases}
 0, r \le R_\mathrm{halo},\\
 M_\mathrm{tot}, r > R_\mathrm{halo}.
\end{cases}
\end{equation}
This model is not physical, but it is extremely valuable for understanding which contributions to gaps will be profile-independent and which effects depend on a specific model's mass distribution. Moreover, it is a building block of eqn.~(\ref{eq:integralFinal}). We find, 
\begin{equation}
    \Delta v_\mathrm{i, shell}(y) \equiv -\frac{2\mathrm{G}M_\mathrm{tot}X_i}{w \rmin^2}\left(1 - \frac{\sqrt{\tilde{R}_\mathrm{max}^2 -  1}}{\tilde{R}_\mathrm{max}} \right)\label{eq:IShell}.
\end{equation}

\subsection{Constant density model}\label{sec:constant}
To further explore varying limits of the internal mass distribution, the extreme limit of a cored density profile is merely a constant density given by 
\begin{equation}
    \rho_\mathrm{constant} = 
    \begin{cases}
\frac{3M_\mathrm{tot}}{4 \pi R_\mathrm{halo}^3}, r \le R_\mathrm{halo},\\
 0, r > R_\mathrm{halo}.
\end{cases}
\end{equation}
Integrating, one finds:
\begin{equation}
    \Delta v_\mathrm{i, constant}(y) \equiv -\frac{2\mathrm{G}M_\mathrm{tot}X_i}{w \rmin^2}\left(1 - \frac{\sqrt{\tilde{R}_\mathrm{max}^2 -  1}}{\tilde{R}_\mathrm{max}}\left( 1- \frac{1}{\tilde{R}_\mathrm{max}^2}\right) \right)\label{eq:IConstant}.
\end{equation}

\subsection{NFW Profile}\label{sec:NFW}
Finally, the NFW profile is defined by \citep{NFW}
\begin{equation}
    \rho = \frac{\rho_0}{(r/r_\mathrm{s})(1+r/r_\mathrm{s})^2},
\end{equation}
where $\rho_0$ is an overall normalization of the density profile and $r_s$ is the scale radius. Integrating, one finds
\begin{equation}
    M_\mathrm{enc}(r) = 4 \pi \rho_0 R_s^3 \left[ \log\left(1 + \frac{r}{r_s}\right) - \frac{r}{r + r_s}\right].\label{eq:NFWMass}
\end{equation}
The pure NFW profile is more dense than the tidally stripped NFW profile. Integrate up to $R_\mathrm{halo, truncated}$, overpredicts gaps by a factor of $\sim 6$. Instead, we invert $M_\mathrm{enc}(r) = M_\mathrm{tot}$ to find
\begin{eqnarray}
R_\mathrm{halo, NFW} &=& r_s\left(1 - \frac{1}{W_0\left[-e^{-\left(1 + \mathcal{M}\right)}\right]} \right), \nonumber \\
\mathcal{M} &\equiv& \frac{M_\mathrm{tot}}{4 \pi \rho_0 r_s^3},\label{eq:lambert}
\end{eqnarray}
where $W_0$ is the principle branch of the Lambert W function \citep{lambert}.  

Therefore, substituting eqn.~(\ref{eq:NFWMass}) into eqn.~(\ref{eq:integralFinalAppendix}), and using eqn.~(\ref{eq:lambert}) as the halo radius, we find
\begin{align}
    \Delta v_\mathrm{i, NFW}(y) = \Delta v_\mathrm{i, shell}(y) \nonumber -  \\ 
    \frac{8\pi G \rho_0 R_s^3X_i}{w\rmin^2}\left(\frac{\sqrt{\tilde{R}^2-1}}{\tilde{R}}\log(a\tilde{R} + 1) - \log(\tilde{R} + \sqrt{\tilde{R}^2 -1})\right. \nonumber + \\
    \left.\left\{\begin{array}{ll}
    \frac{\mathrm{\arctan}\left(\frac{\sqrt{a^2-1}\sqrt{\tilde{R}^2-1}}{a+\tilde{R}}\right)}{\sqrt{a^2-1}}  & \hbox{for } a > 1,\\
    \sqrt{\frac{\tilde{R}-1}{\tilde{R}+1}},  & \hbox{for }  a =1,\\
    \frac{\log\left(\frac{a\tilde{R} +1}{a + \tilde{R} - \sqrt{1-a^2}\sqrt{\tilde{R}^2-1}}\right)}{\sqrt{1-a^2}} & \hbox{for } a < 1,
    \end{array}\right.\right)  \label{eq:NFWKick}
\end{align}
where $a \equiv \rmin/r_\mathrm{s}$ is the inverse normalized scale radius, and  $\tilde{R} = \mathrm{max}\left\{R_\mathrm{halo, NFW}/\rmin, 1\right\}$.  Despite the complicated form, the $a>1$ case is merely the analytic continuation of the $a<1$ where the numerator and denominator become imaginary.  Both cases smoothly approach the $a=1$ case, which can be verified through L'H\^opital's rule.  

This result represents a generalization of \cite{reviewerNFW}'s NFW velocity kick, to include a finite truncation radius. In the untruncated $\tilde{R}\rightarrow \infty$ limit, eqn.~\ref{eq:NFWKick} \emph{nearly} agrees with Table 2 of that work, however there is a sign error in their $s < r_\mathrm{p}$ case, which we have verified with the authors.

\section{Hypergeometric Solution for Tidally Stripped Mass Profile}\label{sec:hypergeometricAppendix}

The tidally stripped mass profile, $M_\mathrm{truncated}(r)$,  cannot be directly integrated to calculate $\mathbf{\Delta v}$. This will be true of any realistic subhalos from any candidate and many other perturbers as well.  Given this general problem, any speed-ups are quite desirable. Solving eqn.~(\ref{eq:integralFinal}) directly takes roughly 24 CPU-hours per timestep, or $\sim 1.5 \times 10^{-2}$ total CPU-hours per realization of Pal-5. While this is quick individually, running $1.6\times 10^6$ realizations of a stream takes $\sim 24,000$ total CPU-hours. For even a single candidate, simulating the $\sim 250$ known streams requires $\sim 6 \times 10^6$ CPU-hours, which grows to completely intractable numbers once several thousand streams have been discovered. Luckily, additional speed-ups are possible, one of which is described (and implemented) below. 

Any continuous mass profile can be represented as power-law splines (piecewise polynomial curves):
\begin{equation}
    M(r)= \alpha_i r^{\beta_i}   \hbox{ for } r_i \leq r < r_{i+1}.\label{eq:powerLawMass}
\end{equation}
For $\sim 30$ outputted radii per subhalo, this spline well approximates NFW-like curves. However, eqn.~\ref{eq:powerLawMass}) is only a linear interpolation of the smoothly varying quantities $(\log(M), \log(r))$. Therefore, only, 17 $(M_i, r_i)$ pairs are outputted for each subhalo, effectively halving the memory requirements of this simulation. Once we have reduced the subhalo population as small as possible using: the filters from Appendix \ref{sec:impactAppendix}, and the conservative requirements $\mathrm{max}(|\Delta v_{y, \mathrm{point}}|) > 0.1$km/s and $\mathrm{min}(\tilde{\rho}_\mathrm{\mathrm{point}}) < 0.9$, we can faithfully reconstruct the other pairs are interpolated with a log-cubic spline, 
\begin{equation}
\log M(r)  = \sum_{j=0}^3 c_{i,j} (\log r)^j \hbox{ for } r_i \leq r < r_{i+1}.
\end{equation}
This ``log-cubic'' spline cannot be used in the techniques that follow, but by employing it to go from $17\rightarrow 30$ $M(r)$ points, we have enough data to represent the curve with eqn.~(\ref{eq:powerLawMass}). Points are chosen non-uniformly to best simulate key regions of this curve.

Now, all that is outputted at this stage is pairs $(M_i, r_i)$ for each subhalo. Therefore, we define
\begin{eqnarray}
    \beta_i = \frac{\log(m_{i+1}/m_i)}{\log(r_{i+1}/r_i)},\label{eq:beta}\\
    \alpha_i = \frac{m_i}{r_i^{\beta_i}},\label{eq:alpha}
\end{eqnarray}
which ensures $\alpha_i r_{i+1}^{\beta_i}= \alpha_{i+1} r_{i+1}^{\beta_{i+1}}$ and $M(r_i) = \alpha_i r_i^{\beta_i}$.
Calling the integral term of eqn.~(\ref{eq:integralFinalAppendix}) $I(\tilde{R}_\mathrm{max})$, we find
\begin{equation}
    I(\tilde{R}_\mathrm{max} ) = \int_1^{\tilde{R}_\mathrm{max}}  \frac{\tilde{M}(\tilde{r})}{ \tilde{r}^2 \sqrt{\tilde{r}^2- 1}}  \mathrm{d}\tilde{r}, 
\end{equation}
or
\begin{equation}
    I(\tilde{R}_\mathrm{max} ) = \sum_{i=1}^N\frac{\alpha_i\rmin^{\beta_i}}{M_\mathrm{tot}}\int_{r_i}^{r_{i+1}}  \frac{ \tilde{r}^{\beta_i -2}}{  \sqrt{\tilde{r}^2- 1}}  \mathrm{d}\tilde{r}, 
\end{equation}
where we have truncated the piecewise mass profile such that $\tilde{r}_1= 1$ and  $\tilde{r}_{N+1} = \tilde{R}_\mathrm{max}$ for convenience. We assume $\tilde{R}_\mathrm{max} >1$ by construction.

Let us redefine
\begin{equation}
    \alpha'_i = \frac{\alpha_i\rmin^{\beta_i}}{M_\mathrm{tot}}.
\end{equation}
Then, since $\beta_i$ are left unchanged by a switch to unitless coordinates,\footnote{This is easily verifiable using eqn.~(\ref{eq:beta}).}  we can rewrite 
\begin{equation}
    I(\tilde{R}_\mathrm{max}) = \sum_{i=1}^N\alpha'_i\int_{r_i}^{r_{i+1}}  \frac{ \tilde{r}^{\beta_i -2}}{  \sqrt{\tilde{r}^2- 1}}  \mathrm{d}\tilde{r}.
\end{equation}
Critically, this integral is solvable using the famous hypergeometric function \citep{hypergeometric}:
\begin{equation}
\begin{split}
    I(\tilde{R}_\mathrm{max}) = \sum_{i=1}^N\frac{i\alpha'_i}{\beta_i-1} \times\\ 
    \left(\tilde{r}_{i+1}^{\beta_i-1}~_{2}F_1\left[1, \frac{\beta_i}{2};\frac{\beta_i+1}{2}; \tilde{r}_{i+1}^2\right] - \tilde{r}_{i}^{\beta_i-1}~_{2}F_1\left[1, \frac{\beta_i}{2};\frac{\beta_i+1}{2}; \tilde{r}_{i}^2\right] \right).\label{eq:hyper1}
\end{split}
\end{equation}
Note that while $ ~_{2}F_1\left[1, \frac{\beta}{2}; \frac{\beta + 1}{2}; r^2\right]$ is complex, the real component of $ r^{\beta-1}~_{2}F_1\left[1, \frac{\beta}{2}; \frac{\beta + 1}{2}; r^2\right]$ is independent of $r$. The differences in eqn.~(\ref{eq:hyper1}) is purely imaginary, and $I(\tilde{R}_\mathrm{max})$ is purely real. Therefore, for any mass profile with non-integrable velocity kicks, we have 

\begin{equation}
    \Delta v_i = \frac{-2\mathrm{G}M_\mathrm{tot}X_i(y)}{w \rmin^2}\left( 1 - \frac{\sqrt{\tilde{R}_\mathrm{max}^2- 1}}{\tilde{R}_\mathrm{max}} + 
    I(\tilde{R}_\mathrm{max}) \right), 
\end{equation}
with 
\begin{equation}
\begin{split}
    I(\tilde{R}_\mathrm{max}) = \sum_{i=1}^N\frac{\alpha'_i}{\beta_i-1}  \times\\ 
    \mathrm{Im}\left(\tilde{r}_{i}^{\beta_i-1}~_{2}F_1\left[1, \frac{\beta_i}{2};\frac{\beta_i+1}{2}; \tilde{r}_{i}^2\right] - \tilde{r}_{i+1}^{\beta_i-1}~_{2}F_1\left[1, \frac{\beta_i}{2};\frac{\beta_i+1}{2}; \tilde{r}_{i+1}^2\right]  \right).\label{eq:hyper2}
    \end{split}
\end{equation}
It is worth repeating that eqn.~(\ref{eq:hyper2}) is not a specific solution to tidally stripped mass profiles, but to any mass profile piecewise-representable by a power law.

This final form takes approximately $1.5$ CPU-hours per timestep to run, or $1500$ CPU-hours in total for a single stream ($20$ CPU-microseconds per subhalo), a $16$ times speed-up. Additional improvements are possible and will be explored in future work.

\bsp	
\label{lastpage}
\end{document}